\documentclass{mn2e}

\usepackage{epsfig}

\renewcommand{\d}{{\rm d}}
\newcommand{\beq}{\begin{equation}}
\newcommand{\eeq}{\end{equation}}
\newcommand{\beqa}{\begin{eqnarray}} 
\newcommand{\eeqa}{\end{eqnarray}}
\newcommand{\bea}{\begin{array}} 
\newcommand{\ea}{\end{array}} 
\newcommand{\lag}{\langle}
\newcommand{\rag}{\rangle}
\newcommand{\Om}{\Omega_{\rm m}}
\newcommand{\Ode}{\Omega_{\rm de}}
\newcommand{\wde}{w_{\rm de}}
\newcommand{\De}{{\cal D}}

\newcommand{\Map}{M_{\rm ap}}

\newcommand{\bk}{{\bf k}}
\newcommand{\kpar}{k_{\parallel}}
\newcommand{\kperp}{\bk_{\perp}}
\newcommand{\kperpDt}{k_{\perp}\De\theta_s}

\newcommand{\wt}{\tilde{w}}

\newcommand{\Mapone}{M_{\rm ap1}}
\newcommand{\Maptwo}{M_{\rm ap2}}
\newcommand{\cM}{{\cal M}}
\newcommand{\cH}{{\cal H}}

\newcommand{\rF}{{\rm F}}

\def\e{\epsilon}

\title[Weak Lensing Surveys]
{Cosmological Constraints from Weak Lensing Surveys}
\author[Munshi \& Valageas]
{Dipak Munshi$^{1,2}$, Patrick Valageas$^{3}$\\
$^{1}$Institute of Astronomy, Madingley Road,
Cambridge, CB3 OHA, United Kingdom\\
$^{2}$Astrophysics Group, Cavendish Laboratory, Madingley Road, 
Cambridge CB3 OHE, United Kingdom\\
$^{3}$Service de Physique Th\'eorique, 
CEA Saclay, 91191 Gif-sur-Yvette, France \\}

\begin{document}
\maketitle

\begin{abstract}
Focusing on the well motivated aperture mass statistics $\Map$, we study the 
possibility of constraining cosmological parameters using future space based 
SNAP class weak lensing missions. Using completely analytical results we 
construct the covariance matrix for estimators based on two-point and 
three-point statistics. Our approach incorporates an accurate modelling of 
higher-order statistics to describe cosmic variance as well as various 
sources of discrete noise at small angular scales. These results are then fed
into a Fisher matrix based analysis to study cosmological parameter 
degeneracies. Joint and independent analysis, with or without redshift 
binning, for various parameter combinations are presented. An analytical 
modelling of the covariance matrix opens up the possibility of testing 
various approximations which are often used in derivations of semi-analytical 
results. These include how inclusion of full non-Gaussian terms in covariance 
matrix affects parameter estimation. Inclusion of three-point information and 
how such information can enhance the accuracy with which certain parameters 
can be estimated is also studied in detail. It is shown that broad correlation 
structure among various angular scales in such circumstances mean reduction 
in number of available angular scales which carry completely independent 
information. On the other hand, the effect of theoretical inaccuracies, in 
modelling either the power-spectrum or bi-spectrum evolution, onto the measure 
of cosmological parameters from weak lensing surveys is also considered.
Several cosmological parameters, $\Om$, $\sigma_8$, spectral index $n_s$, 
running of spectral index $\alpha_s$ and equation of state of the dark energy 
$\wde$ are included in the analysis.
\end{abstract}


\section{Introduction}
\label{Introduction}

Weak lensing surveys are expected to make significant contribution in 
constraining cosmology by studying the statistics of dark matter distribution 
in the nearby universe in an unbiased way. For instance, Contaldi et al.(2003) 
used the Red Cluster Sequence (RCS) to show that the $\Om$ and $\sigma_8$ 
degeneracy directions are nearly orthogonal making them particularly suitable 
for combined analysis (van Waerbeke et al. 2002). Ishak et al. (2003) 
argued that joint CMB-cosmic shear surveys provide an optimal data set for 
constraining the amplitude and running of spectral index which helps to probe 
various inflationary models. Tereno et. al.(2004) have studied cosmological 
forecasts for joint CMB and weak lensing survey data. Clearly, the potential 
of weak lensing  surveys (Mellier 1999; Bartelmann \& Schneider 2001; 
R\'efr\'egier 2003; van Waerbeke \& Mellier 2003) as a cosmological probe is 
now well established (Contaldi et al. 2003; Hu \& Tegmark 1999). In last few 
years there have been many studies which have directly detected cosmological 
shear in random patches of the sky (Brown et al. 2003;
Bacon et al. 2002; Bacon, Refregier \& Ellis 2000; Hamana et
al. 2003; H\"{a}mmerle et al. 2002; Hoekstra et al. 2002a; Hoekstra,
Yee \& Gladders 2002a; Jarvis et al.  2002; Kaiser, Wilson
\& Luppino 2000; Maoli et al. 2001; Refregier, Rhodes, \& Groth 2002;
Rhodes, Refregier \& Groth 2001; van Waerbeke et al. 2000; van
Waerbeke et al. 2001a; van Waerbeke et al. 2002; Wittman et
al. 2000). While early studies were primarily concerned with detection of
weak lensing signal, present generation of weak lensing studies are putting
constraints on cosmological parameters such as matter density parameter 
$\Om$ and power spectrum normalisation $\sigma_8$. Not only these studies can 
put remarkable constraints on cosmological parameters, they can help to break 
up parameter degeneracies when used along with other cosmological surveys
such as CMB observations as we discussed above.

% Future surveys

Inspired by the success of these surveys, there are many other ongoing, 
planned and proposed weak lensing surveys which are currently in progress
such as the Deep Lens Survey (Wittman et al. 2002), the NOAO Deep survey 
(Mellier et al. 2001) \footnote{http://www.noao.edu/noao/noaodeep/}, the 
Canada-France-Hawaii Telescope Legacy Survey (Mellier et al. 2001) 
\footnote{http://www.cfht.hawaii.edu/Science/CFHLS/}, the Panoramaic Survey 
Telescope and Rapid Response System 
\footnote{http://pan-starrs.ifa.hawaii.edu/public/},
the Supernova Acceleration Probe \footnote{http://snap.lbl.gov/} (Rhode et. al. 2003; Massey et al. 2003) and the large Synoptic Survey Telescope \footnote{http://www.lsst.org/} (Tyson et. al. 2002). Independently or jointly these 
observations can contribute to increase the accuracy with which various 
cosmological parameters can be determined from other cosmological observation 
such as CMB. While first generations of cosmic shear surveys have demonstrated 
(van Waerbeke \& Mellier 2003; Refregier 2003) the feasibility of weak lensing 
studies in constraining  the dark matter power spectrum parametrised by 
$\sigma_8$, $\Om$ and shape parameter $\Gamma$ (Munshi \& Valageas 2005 
contains a recent compilations of present ground based survey results), the 
future surveys will be able to probe much larger scales and therefore it will 
be possible to study the linear regime directly to put more stringent bounds 
on cosmological parameters such as the equation of state of dark energy and 
its time variations.

% Motivation

Various groups have carried out detailed analysis of cosmological constraints 
that are achievable from ground and space based weak lensing surveys 
(Bernardeau et al. 1997, Jain \& Seljak 1997, Kaiser 1998). Early studies 
focused mainly on a limited number of cosmological parameters
whereas later on it was realised that weak lensing can play an important role 
in constraining the dark energy equation of state (Hu 1999, 2002ab; 
Hueter 2002; Abazajjian \& Dodelson 2003; Heavens 2003; Refrefier et al. 2003; 
Benabed \& van Waerbeke 2003; Jain \& Taylor 2003 and references therein). 
Hu \& Tegmark (2004) studied a general multi-parameter Fisher matrix analysis 
based on power spectrum to check prospects of weak lensing for parameter 
estimation independently or jointly with external data set such as CMB. 
Takada \& Jain (2005) have presented a study based on joint analysis of power 
spectrum and bi-spectrum and emphasised the role of tomography in in weak 
lensing parameter estimation. More recently Kilbinger \& Schneider (2005) have 
studied joint constraints from second- and third- order
moments based on aperture mass $\Map$ statistics to study cosmological 
parameter estimation efficiency. Although similar in motivation we use 
a different analytical approach to check the efficiency of future space based 
weak lensing surveys to constrain cosmology. We combine measurements from 
second $\lag\Map^2\rag$ and third order $\lag\Map^3\rag$ statistics to 
constrain cosmological parameters such as $\Om$, $\sigma_8$, $\alpha_s$, $n_s$ 
and $\wde$. The effect of estimating the median redshif $z_0$ of the source 
redshift distribution directly from the data is also included. Impact of 
inaccurate modelling of power-spectrum and bi-spectrum of the
underlying mass distribution is also analysed.

The reasons for choosing $\Map$ over other two-point statistics, as pointed 
out by Kilbinger \& Schneider (2005), are several. Unlike other two-point 
statistics $\Map$  by construction is a scalar object and unlike many other 
three-point objects, constructed from shear three-point correlation functions,
the third order moment of $\Map$ has a non-vanishing ensemble average and 
hence can directly probe the underlying bi-spectrum of the matter distribution.
Besides, $\Map$ and its extensions has the inherent characteristic of 
separating gravity induced $E$ mode from $B$ mode (see e.g. Crittenden et al. 
2001, for more detailed discussion on this issue) generated mainly by residual
systematics which can affect both second- and third-order estimators. 
Moreover, it is to be noted that the integral measure $\lag\Map^3\rag$ is also 
a local measure of the bi-spectrum and does not suffer from the oscillatory 
nature of other filters.

This paper is organised as follows. In section~2 we introduce the second- and 
third- order estimators based on $\Map$ statistics. Expressions are presented 
for their joint covariances. Section~3 describes the cosmological parameters 
and the survey design for proposed SNAP class experiments.
Next, in sections 4 and 5 we study the influence of non-Gaussianities onto
the covariance matrices and the signal to noise ratios.
In section~6 we introduce the Fisher matrix formalism in our context 
and use it to study estimation error for weak lensing observable. Finally the 
section~5 is left for discussion of our results and future prospects. 
Appendix~A is devoted to comparing the performance of ground based surveys
and SNAP class space based surveys. It also takes a detailed look at various 
issues of weak lensing survey design.

\section{Aperture-mass statistics}
\label{Aperture-mass}

\subsection{Weak-lensing effects}
\label{Weak-lensing}

Weak gravitational lensing effects probe the matter distribution integrated
along the line of sight to distant sources like galaxies (e.g., Bernardeau 
et al. 1997; Kaiser 1998). Therefore, they
can be used to obtain information on the 2-D projected density field $\kappa$
(also called the convergence field) on the sky. In practice, one needs to take
into account the redshift distribution $n(z_s)$ of the sources. Indeed, in 
order to have good statistics one needs to average weak lensing effects over
many sources which entails rather broad source distributions. On the other 
hand, it can be convenient to use a filtered version of the convergence 
$\kappa$, such as the aperture-mass $\Map$. More specifically, the latter is 
obtained from $\kappa$ by using a compensated filter. Then, one can also
express $\Map$ as a function of the tangential shear $\gamma_{\rm t}$ 
(Kaiser 1994; Schneider 1996) which can be directly estimated from the observed
ellipticities of distant galaxies. This is the property which makes the
aperture-mass a useful quantity. In addition, since it involves a compensated 
filter the contribution from long wavelengths to the signal is damped as 
compared with a top-hat filter (which would yield the mean convergence) so
that $\Map$ allows one to probe the matter density field over a narrow range 
of wavelengths which can be varied through the angular scale $\theta_s$ of the 
window $U_{\Map}({\vec \vartheta})$. Thus, the aperture-mass can be written
in terms of the fluctuations of the density field as:
\beq
\Map= \int\d{\vec \vartheta}\; U_{\Map}({\vec \vartheta}) \; 
\int_0^{\chi_{\rm max}} \d\chi \; \wt(\chi) \; 
\delta(\chi,\De{\vec \vartheta}) , 
\label{Mapdelta}
\eeq
with:
\beq 
\wt(\chi) = \frac{3\Om}{2} \int_z^{z_{\rm max}} \d z_s \; n(z_s) \;
\frac{H_0^2}{c^2} \; \frac{\De(\chi) \De(\chi_s-\chi)}{\De(\chi_s)} \; (1+z) .
\label{wt}
\eeq
Here the redshift $z$ corresponds to the radial distance $\chi$ and $\De$ is 
the angular distance, ${\vec \vartheta}$ is the angular direction on the sky, 
$\delta(\chi,\De{\vec \vartheta})$ is the matter density contrast and 
hereafter we normalise the mean redshift distribution of the sources 
(e.g. galaxies) to unity: $\int\d z_s \; n(z_s)=1$. 
We defined $z_{\rm max}$ as the depth of the survey
(i.e. $n(z_s)=0$ for $z_s>z_{\rm max}$). Here and in the following we use the 
Born approximation which is well-suited to weak-lensing studies: the 
fluctuations of the gravitational potential are computed along the 
unperturbed trajectory of the photon (Kaiser 1992). We also neglect the 
discrete effects due to the finite number of galaxies. They can be obtained 
by taking into account the discrete nature of the distribution $n(z_s)$. 
This gives corrections of order $1/N$ to higher-order moments of weak-lensing 
observable, where $N$ is the number of galaxies within the circular field 
of interest. In practice $N$ is much larger than unity (for a circular 
window of radius 1 arcmin we expect $N \ga 100$ for the SNAP mission) 
therefore in this paper we shall work with eq.(\ref{Mapdelta}). 
We choose for the filter $U_{\Map}$ associated with the aperture-mass $\Map$ 
the window of finite support used in Schneider (1996):
\beq
U_{\Map} = \frac{\Theta(\vartheta<\theta_s)}{\pi\theta_s^2}
\; 9 \left(1-\frac{\vartheta^2}{\theta_s^2}\right) 
\left(\frac{1}{3} - \frac{\vartheta^2}{\theta_s^2}\right) ,
\label{UMap}
\eeq
where $\Theta$ is a top-hat with obvious notations. The angular radius 
$\theta_s$ gives the angular scale probed by this smoothed observable.
As described in Munshi et al. (2004), the cumulants of $\Map$ can be
written in Fourier space as:
\beqa
\lag \Map^p \rag_c & = & \int_0^{\chi_{\rm max}} \prod_{i=1}^{p} \d\chi_i \; 
\wt(\chi_i) \int \prod_{j=1}^{p} \d\bk_j \; 
W_{\Map}(\bk_{\perp j} \De_j \theta_s) \nonumber \\
& & \times \; \left( \prod_{l=1}^{p} e^{i k_{\parallel l} \chi_l} \right) 
\;\; \lag \delta(\bk_1) \dots \delta(\bk_p) \rag_c .
\label{cumMapk}
\eeqa
We note $\lag .. \rag$ the average over different realizations of the
density field, $\kpar$ is the component of $\bk$ parallel to the 
line-of-sight, $\kperp$ is the two-dimensional vector formed by the 
components of $\bk$ perpendicular to the line-of-sight and 
$W_{\Map}(\kperp\De\theta_s)$ is the Fourier transform of the window 
$U_{\Map}$:
\beq
W_{\Map}(\kperp\De\theta_s) = \int\d{\vec \vartheta} \; 
U_{\Map}({\vec \vartheta}) \; e^{i \kperp.\De{\vec \vartheta}} =  
\frac{24 J_4(\kperpDt)}{(\kperpDt)^2} .
\label{WMap}
\eeq
The Fourier-space expression (\ref{cumMapk}) is well suited to models which 
give a simple expression for the correlations 
$\lag \delta(\bk_1) .. \delta(\bk_p)\rag_c$, such as the stellar model 
(Valageas et al. 2004; Barber et al. 2004) defined by:
\beq
\lag \delta(\bk_1) .. \delta(\bk_p)\rag_c = \frac{\tilde{S}_p}{p} \; 
\delta_D(\bk_1+\dots+\bk_p) \; \sum_{i=1}^p \prod_{j\neq i} P(k_j) , 
\label{stellar}
\eeq
where $\tilde{S}_2=1$, $\delta_D$ is the Dirac distribution and $P(k)$ is the 
3-D power-spectrum of the density fluctuations. The coefficients $\tilde{S}_3,
\tilde{S}_4,..$ are closely related (and approximately equal) to the skewness,
kurtosis, .., of the density field. Eq.(\ref{cumMapk}) generalises in a
straightforward fashion for many-point cumulants. This allows one to consider
for instance the cross-correlations between the two statistics $\Mapone$ and
$\Maptwo$ associated with two different angular scales 
$\theta_{s1},\theta_{s2}$ and/or two different redshift distributions 
$n_1(z_s),n_2(z_s)$.

\subsection{Low-order estimators}
\label{Low-order}

The expressions recalled in the previous section describe weak lensing effects
due to the fluctuations of the matter density field. In practice, one actually
measures the aperture-mass $\Map$ from the distortions of the images of 
distant sources. Thus, in the case of weak lensing ($|\kappa| \ll 1$) the 
observed complex ellipticity $\epsilon=\epsilon_1+i\epsilon_2$ of a distant 
galaxy is related to the shear $\gamma=\gamma_1+i\gamma_2$ by: 
$\epsilon=\gamma+\epsilon_*$, where $\epsilon_*$ is the intrinsic ellipticity 
of the galaxy. On the other hand, the aperture-mass defined in 
eqs.(\ref{Mapdelta})-(\ref{UMap}) can also be written as 
a function of the tangential shear $\gamma_{\rm t}$ as (Kaiser et al. 1994, 
Schneider 1996) as:
\beq
\Map= \int \d{\vec \vartheta} \; Q_{\Map}({\vec \vartheta}) \; 
\gamma_{\rm t}({\vec \vartheta})
\label{Mapgammat}
\eeq
with:
\beq
Q_{\Map}({\vec \vartheta}) = \frac{\Theta(\vartheta<\theta_s)}{\pi\theta_s^2}
\; 6  \; \left(\frac{\vartheta}{\theta_s}\right)^2 
\left(1-\frac{\vartheta^2}{\theta_s^2}\right) .
\label{QMap}
\eeq
This leads us to define the estimators $M_p$ for low-order moments (e.g.,
Munshi \& Coles 2003, Valageas et al. 2004b):
\beq
M_p = \frac{(\pi \theta_s^2)^p} {(N)_p} \sum_{(j_1,\dots,j_p)}^N 
Q_{j_1} \dots Q_{j_p} \; \e_{{\rm t}j_1} \dots \e_{{\rm t}j_p}, 
\label{Mp}
\eeq
with:
\beq
(N)_p = N (N-1) .. (N-p+1) = \frac{N!}{(N-p)!} ,
\eeq
where $N$ is the number of galaxies in the patch of size $\pi\theta_s^2$, 
$p$ is the order of the moment and $Q_j=Q_{\Map}({\vec \vartheta}_j)$ where
${\vec \vartheta}_j$ and $\epsilon_{{\rm t}j}$ are the direction on the sky 
and the observed tangential ellipticity of the galaxy $j$.
Finally, the sum in eq.(\ref{Mp}) runs over 
all sets of $p$ different galaxies among the $N$ galaxies enclosed 
in the angular radius $\theta_s$, which ensures that 
$\lag M_p\rag=\lag \Map^p \rag$ if we neglect the correlation of the intrinsic
ellipticity of a given galaxy with other galaxies or with weak lensing 
observables. 

These estimators $M_p$ correspond to a single circular field of angular 
radius $\theta_s$ containing $N$ galaxies. In practice, the size of the survey 
is much larger than $\theta_s$ and we can average over $N_c$ cells on the sky. 
This yields the estimators $\cM_p$ defined by:
\beq
\cM_p = \frac{1}{N_c} \sum_{n=1}^{N_c} M_p^{(n)} , \;\;\; \mbox{whence} \;\;\;
\lag\cM_p\rag = \lag \Map^p \rag ,
\label{cMp}
\eeq
where $M_p^{(n)}$ is the estimator $M_p$ for the cell $n$ and we assumed 
that these cells are sufficiently well separated so as to be uncorrelated.
The estimators $M_p$ and $\cM_p$ provide a measure of the moments 
$\lag \Map^p\rag$, which can be used to constrain cosmological parameters.
However, as shown in Valageas et al. (2004b), it is better to first consider 
cumulant-inspired estimators $H_p$. Thus, for the second and third-order 
cumulants we define:
\beq
H_2 = M_2 , \;\;\;  H_3 = M_3 - 3 \cM_2 M_1 ,
\label{H3}
\eeq
and:
\beq
\cH_p = \frac{1}{N_c} \sum_{n=1}^{N_c} H_p^{(n)} , \;\;\; \lag\cH_2\rag = 
\lag\Map^2\rag_c , \;\;\; \lag \cH_3 \rag = \lag \Map^3 \rag_c .
\label{cHp}
\eeq
The interest of $\cH_3$  is that its scatter is smaller than for
$\cM_3$, see Valageas et al.(2004b). Besides it directly yields the
one-point cumulants (here we neglected higher-order terms over $1/N_c$).

\subsection{Covariance matrix}
\label{Covariance}

The estimators $\cM_p$ and $\cH_p$ defined in the previous section allow one
to estimate the cumulants of the aperture-mass $\lag\Map^p\rag_c$ for
a given angular scale $\theta_s$ and source redshift distribution $n(z_s)$.
In practice we can vary both the angular scale $\theta_s$ of the filters
$U_{\Map}$ and $Q_{\Map}$ and the redshift distribution $n(z_s)$ (for instance
by applying a simple binning of the galaxies over redshift and selecting 
different redshift bins, which is often referred to as tomography). Thus,
if we restrict ourselves to second-order moments $\lag\Map^2\rag$ we obtain the
set of estimators $\cH_2(i)$, with $\lag\cH_2(i)\rag=\lag\Map^2(i)\rag$ as 
defined in eq.(\ref{cHp}), where the index $i=1,..,N_i$ stands for both the
angular scale $\theta_{si}$ and the redshift distribution $n_i(z_s)$ 
(for $N_{\theta}$ angular scales and $N_z$ redshift bins we have 
$N_i=N_{\theta}\times N_z$).
Then, the covariance matrix $C_{ij}$ associated with this data set is:
\beq
C_{ij} = \lag\cH_2(i)\cH_2(j)\rag - \lag\cH_2(i)\rag \lag\cH_2(j)\rag .
\label{Cij}
\eeq
It measures the cross-correlation between the two estimators $\cH_2(i)$ and 
$\cH_2(j)$. In the following, we shall consider the limit where the number of
cells $N_c$ on the sky goes to infinity, that is the sum in eq.(\ref{cHp})
is performed over $N_c$ cells which cover the whole survey area with a uniform
coverage and which are separated by an angular shift $\Delta{\vec\alpha}$ 
which goes to zero. Therefore, the discrete sum (\ref{cHp}) tends to the
integral:
\beq
\cH_2(i) = \int_A \frac{\d{\vec\alpha}}{A} \; H_2(i;{\vec\alpha}) ,
\label{cH2i}
\eeq
where $A$ is the survey area and $H_2(i;{\vec\alpha})$ is the estimator 
$H_2(i)$, associated with the angular scale $\theta_{si}$ and redshift 
distribution $n_i(z_s)$, for the cell centred on the direction ${\vec\alpha}$
on the sky. Hereafter we neglect side effects and we consider that all 
estimators cover the same area $A$, independently of the radius $\theta_{si}$.
Then, after one angular integration we obtain for the covariance $C_{ij}$:
\beq
C_{ij} = \int_A \frac{\d{\vec\alpha}}{A} \; 
\sigma^2(H_2(i),H_2(j);{\vec\alpha}) ,
\label{Cijalpha}
\eeq
where $\sigma^2(H_2(i),H_2(j);{\vec\alpha})$ is the 
cross-correlation between the two estimators $H_2(i)$ and $H_2(j)$ 
separated by the angular shift ${\vec\alpha}$. From eq.(\ref{Mp}) the
cross-correlation $\sigma^2$ reads (see also the expressions given in App.A1 
of Munshi \& Valageas 2005):
\beqa
\lefteqn{ \sigma^2(H_2(i),H_2(j);{\vec\alpha}) =
\lag\Map^2(i)\Map^2(j)\rag_c({\vec\alpha}) } \nonumber \\
&& + 2 [ \lag\Map(i)\Map(j)\rag_c({\vec\alpha}) + Q_{ij}({\vec\alpha}) ]^2 ,
\label{sigma2}
\eeqa
where we recalled explicitly the dependence on the angular shift 
${\vec\alpha}$, and we introduced $Q_{ij}$ defined by:
\beq
Q_{ij}({\vec\alpha}) = \frac{\sigma_*^2}{2} \frac{n_{{\rm g}ij}}
{n_{{\rm g}i}n_{{\rm g}j}} \int \d{\vec\vartheta} \; Q_i({\vec\vartheta}) 
Q_j({\vec\vartheta} - {\vec\alpha}) ,
\label{Qij}
\eeq
where $\sigma_*^2=\lag|\epsilon_*|^2\rag$ is the scatter of the galaxy 
intrinsic ellipticities, $n_{{\rm g}i}$ is the number surface density of 
galaxies associated with the redshift distribution $n_i(z_s)$,
$n_{{\rm g}ij}$ is the surface density of common galaxies to both redshift 
distributions $n_i(z_s)$ and $n_j(z_s)$, and $Q_i$ is the filter $Q_{\Map}$
defined in eq.(\ref{QMap}) for the radius $\theta_{si}$. We can obtain
in a similar fashion the covariance matrix associated with the data set 
$\{\cH_3(i)\}$, associated with third-order cumulants, as well as the full
data set $\{\cH_2(i),\cH_3(i)\}$ where we consider both second-order and
third-order cumulants. Thus, we have (see also Munshi \& Valageas 2005):
\beqa
\lefteqn{\sigma^2(H_2(i),H_3(j)) = \lag\Map^2(i)\Map^3(j)\rag_c }
\nonumber \\
&&  + 6 \lag\Map(i)\Map^2(j)\rag_c [ \lag\Map(i)\Map(j)\rag_c + Q_{ij} ] ,
\label{sigma23}
\eeqa
and:
\beqa
\lefteqn{\sigma^2(H_3(i),H_3(j)) = \lag\Map^3(i)\Map^3(j)\rag_c } \nonumber \\
&& + 9 \lag\Map^2(i)\Map^2(j)\rag_c [ \lag\Map(i)\Map(j)\rag_c + Q_{ij} ] 
\nonumber \\
&& + 9 \lag\Map^2(i)\Map(j)\rag_c \lag\Map(i)\Map^2(j)\rag_c \nonumber \\
&& + 6 [ \lag\Map(i)\Map(j)\rag_c + Q_{ij} ]^3 ,
\label{sigma33}
\eeqa
where we did not recall explicitly the dependence on the angular shift 
${\vec\alpha}$.

It is interesting to estimate the amplitude of various contributions to the
covariance matrix $C_{ij}$, due to noise, Gaussian and non-Gaussian terms.
Thus, in addition to the full matrix $C_{ij}$ described above we also
introduce the matrices $C_{ij}^{G,c.v.}$, $C_{ij}^{G,s.n.}$ and 
$C_{ij}^{c.v.}$. The matrix $C_{ij}^{G,c.v.}$ only includes Gaussian terms 
which contribute to the cosmic variance (i.e. non-Gaussianities and the noise 
are set to zero). This yields for instance for 
$\sigma^2(H_2(i),H_2(j))^{G,c.v.}$:
\beq
\sigma^2(H_2(i),H_2(j))^{G,c.v.} = 2 \lag\Map(i)\Map(j)\rag_c^2 .
\label{sigma2Gcv}
\eeq
We obtain from eq.(\ref{sigma33}) a similar expression for 
$\sigma^2(H_3(i),H_3(j))^{G,c.v.}$, while from eq.(\ref{sigma23}) we see 
at once that $\sigma^2(H_2(i),H_3(j))^{G,c.v.}=0$. Next, the matrix 
$C_{ij}^{G,s.n.}$ includes all Gaussian terms which involve the shot noise. 
This yields for $\sigma^2(H_2(i),H_2(j))^{G,s.n.}$:
\beq
\sigma^2(H_2(i),H_2(j))^{G,s.n.} = 4 \lag\Map(i)\Map(j)\rag_c Q_{ij} +
2 Q_{ij}^2 .
\label{sigma2Gsn}
\eeq
We again obtain from eq.(\ref{sigma33}) a similar relation for 
$\sigma^2(H_3(i),H_3(j))^{G,s.n.}$, while from eq.(\ref{sigma23}) we see 
at once that $\sigma^2(H_2(i),H_3(j))^{G,s.n.}=0$. Note that 
$C_{ij}^{G,c.v.}+C_{ij}^{G,s.n.}$ is equal to the matrix $C_{ij}$
without non-Gaussian terms. Finally, the matrix $C_{ij}^{c.v.}$ is equal to
the matrix $C_{ij}$ where we set the noise equal to zero. It includes both
Gaussian and non-Gaussian terms, and we obtain for 
$\sigma^2(H_2(i),H_2(j))^{c.v.}$:
\beq
\sigma^2(H_2(i),H_2(j))^{c.v.} = \lag\Map^2(i)\Map^2(j)\rag_c
+ 2 \lag\Map(i)\Map(j)\rag_c^2 .
\label{sigma2cv}
\eeq
From eqs.(\ref{sigma23})-(\ref{sigma33}) we obtain similar expressions for
$\sigma^2(H_2(i),H_3(j))^{c.v.}$ and $\sigma^2(H_3(i),H_3(j))^{c.v.}$.
Note that $\sigma^2(H_2(i),H_3(j))^{c.v.}$ is different from zero.
Thus, the comparison between the four matrices $C_{ij}$, 
$C_{ij}^{G,c.v.}$, $C_{ij}^{G,s.n.}$ and $C_{ij}^{c.v.}$
allows us to evaluate the relative importance of the shot noise (associated
with the dispersion $\sigma_*^2$ of the galaxy intrinsic ellipticity) and of
the cosmic variance (merely due to the finite size of the survey), as well as
the relative importance of Gaussian and non-Gaussian terms.

\section{Background cosmology and survey parameters}
\label{Background}

We describe in this section the background cosmology and the specific 
correlation hierarchy for the matter distribution which we use in numerical
computations. We also focus on the SNAP observational strategy to estimate
the accuracy which can be obtained from such weak lensing surveys.
However, as mentioned before the basic formalism 
developed here remains completely general and specific details studied here 
only serve illustrational purposes.

\subsection{Cosmological parameters}
\label{Cosmological}

For the background cosmology we consider a fiducial $\Lambda$CDM model with
$\Om=0.3$, $\Ode=0.7$, $\wde=-1$, $H_0=70$ km/s/Mpc and $\sigma_8=0.88$.
We note $\Om$ the matter density (cold dark matter plus baryons) 
and $\Ode$ the dark-energy density today at $z=0$. We parameterize its
equation of state by $\wde=p_{\rm de}/\rho_{\rm de}$. For $\wde=-1$
this corresponds to a simple cosmological constant. In the following we shall
investigate the dependence of weak-lensing observable on the cosmological
parameters $\Om$ (keeping $\Om+\Ode=1$ for a flat universe) and $\wde$.
We always consider $\wde$ to be a constant independent of time. Then, the
Hubble expansion rate $H(z)$ reads (e.g., Linder \& Jenkins 2003):
\beq
\frac{\dot{a}}{a} = H(z)= H_0 \sqrt{\Om(1+z)^3+\Ode(1+z)^{3(1+\wde)}} ,
\label{Hz}
\eeq
where $a(t)=(1+z)^{-1}$ is the cosmological scale factor while the dot 
denotes the derivative with respect to physical time (we used $\Om+\Ode=1$). 
The linear growth factor $D(t)$ obeys the differential equation:
\beq
\ddot{D} + 2 H \dot{D} - \frac{3}{2} \Om H_0^2 (1+z)^3 D = 0 ,
\label{Dt}
\eeq
which yields for the linear growth rate $g(a)$ relative to a critical-density
universe $g=D/a$:
\beqa
\lefteqn{ 2 \frac{\d^2 g}{\d\ln a^2} + \frac{5\Om+(5-3\wde)\Ode a^{-3\wde}}
{\Om+\Ode a^{-3\wde}} \frac{\d g}{\d\ln a} } \nonumber \\ 
&& + \frac{(3-3\wde)\Ode a^{-3\wde}}{\Om+\Ode a^{-3\wde}} g = 0 .
\label{ga}
\eeqa
We obtain $g(a)$ by solving numerically the differential eq.(\ref{ga}).

For the matter transfer function $T(k)$ we use the fitting formulae provided 
by Eisenstein \& Hu (1999), with $\Omega_{\rm b}=0.047$, $n_s=1$ and 
$\alpha_s=0$. Here $n_s$ is the power-law index of the primordial 
power-spectrum and $\alpha_s$ is the running index. More precisely, we write 
the linear matter power-spectrum as in Spergel et al. (2003):
\beq
P_L(k) \propto \left(\frac{k}{k_0}\right)^{n_s+\alpha_s \ln(k/k_0)/2} T^2(k),
\label{PLk}
\eeq
where $k_0=0.05$ Mpc$^{-1}$. Thus, the local slope $n$ of the primordial 
spectrum is:
\beq
n=\frac{\d\ln P_{\rm prim}}{\d\ln k} = n_s+\alpha_s \ln
\left(\frac{k}{k_0}\right) , \;\;\; \frac{\d n}{\d\ln k} = \alpha_s .
\label{nsPk}
\eeq
Next, we use the fit given by Peacock and Dodds (1996) to model the non-linear 
evolution of the matter power-spectrum due to gravitational clustering.
In order to investigate the sensitivity of weak lensing data on this 
non-linear extrapolation we introduce an additional parameter $f_2$. Thus, we
modify eq.(21) of Peacock \& Dodds (1996) as:
\beq
f_{\rm NL}(x) = x \left[ \frac{1+B\beta x+[A x]^{\alpha\beta}} 
{1+([A x]^{\alpha}g^3(\Omega_m)/[f_2 V x^{1/2}])^{\beta}} \right]^{1/\beta} . 
\label{f2}
\eeq
This is identical to the formula from Peacock \& Dodds (1996) for $f_2=1$ (see
their paper for the meaning of various parameters). In the linear regime we 
merely have $f_{\rm NL}(x)= x$ while in the non-linear regime we have
$f_{\rm NL}(x) \propto f_2 x^{3/2}$. Therefore, the parameter $f_2$ describes
the amplitude of the non-linear part of the matter power-spectrum as compared 
with Peacock \& Dodds (1996) (which corresponds to $f_2=1$).

\begin{figure}
\begin{center}
\epsfxsize=4.1 cm \epsfysize=5 cm {\epsfbox{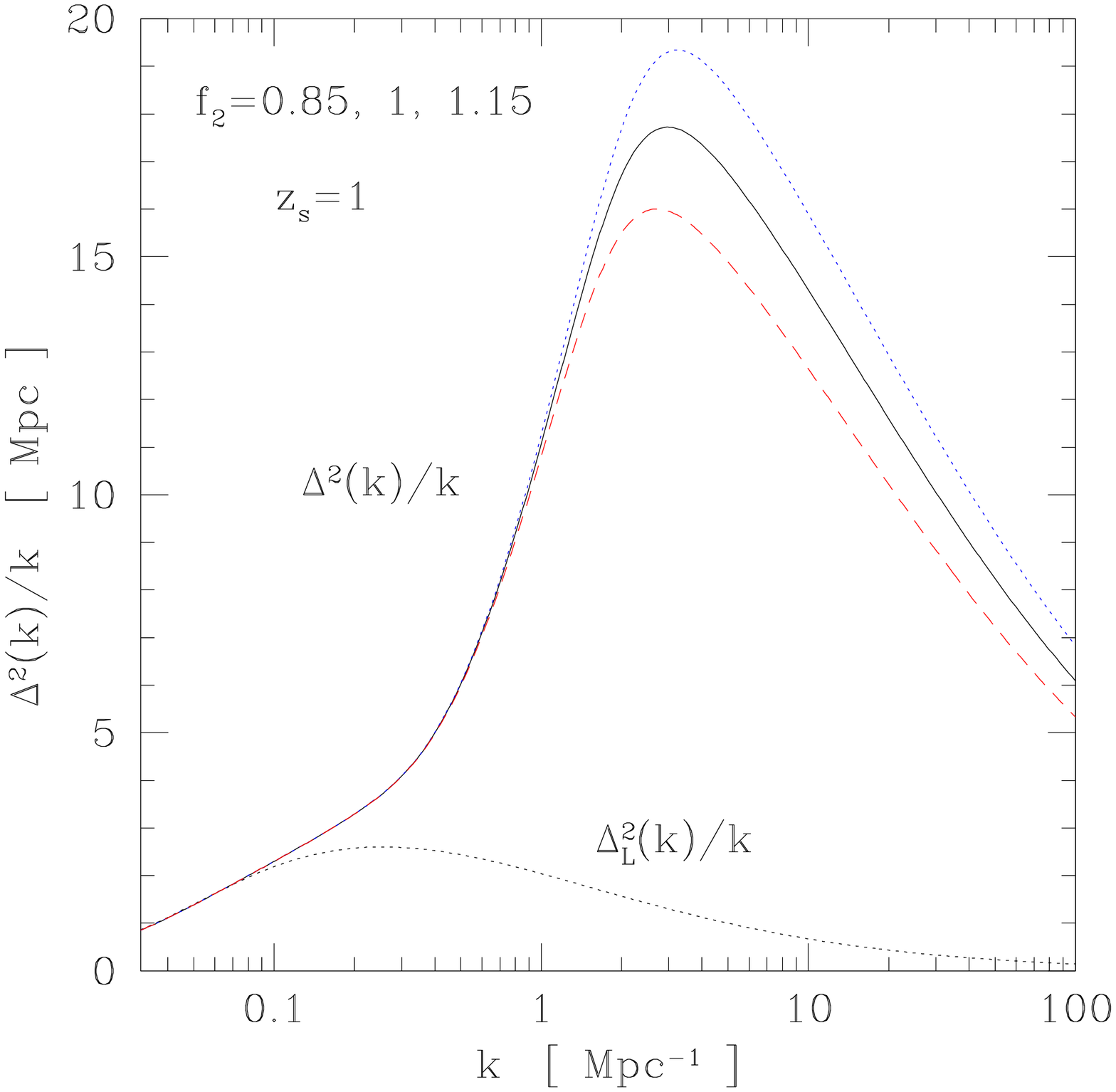}} 
\epsfxsize=4.2 cm \epsfysize=5 cm {\epsfbox{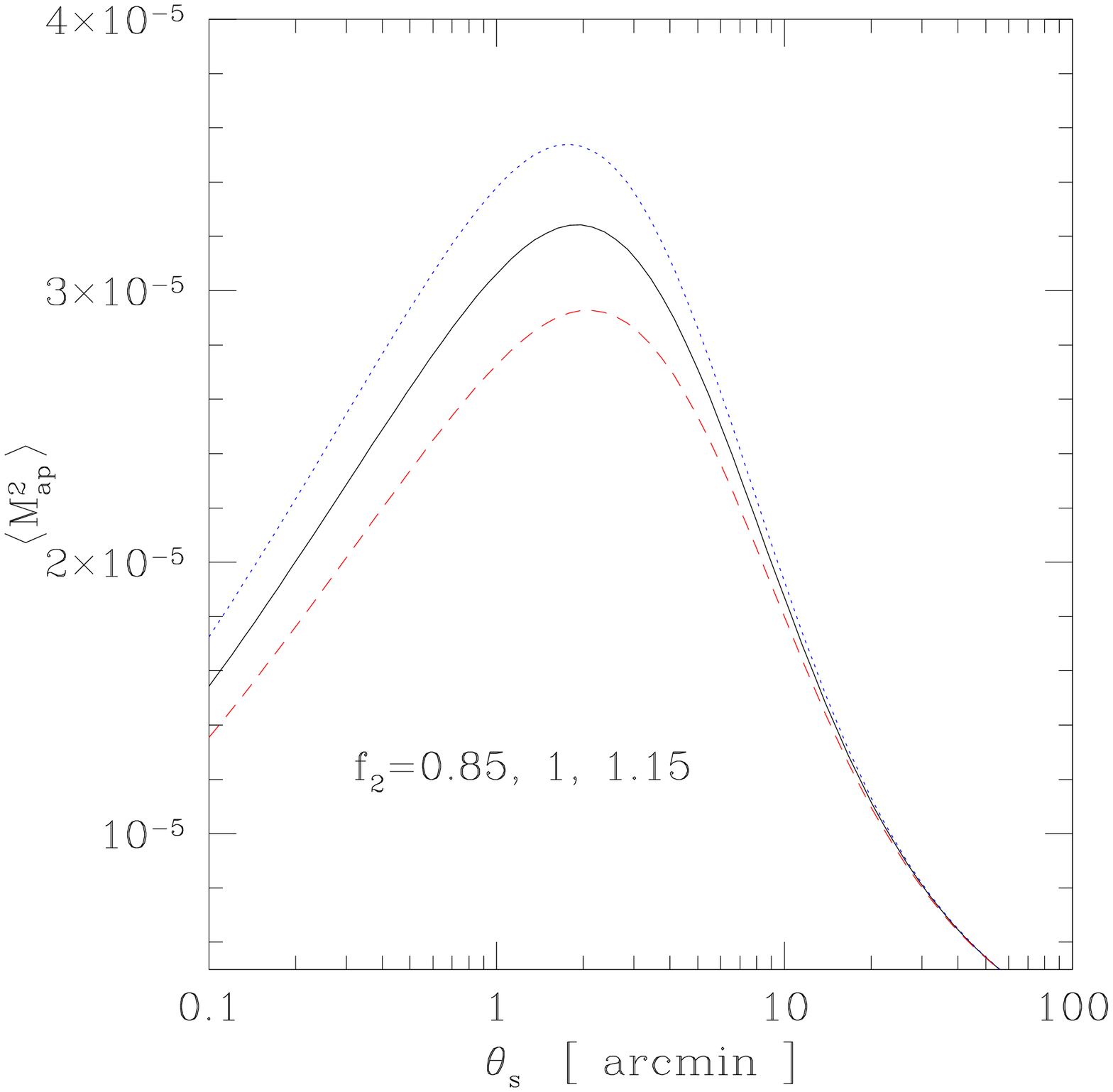}}
\end{center}
\caption{{\it Left panel:} The power $\Delta^2(k)/k$ at redshift $z=1$ as a 
function of the comoving wavenumber $k$. The three upper curves correspond to 
$f_2=0.85$ (lower dashed curve), $f_2=1$ (middle solid line, this corresponds 
to Peacock \& Dodds 1996) and $f_2=1.15$ (upper dotted curve). This 
corresponds roughly to a $10\%$ variation of the matter power-spectrum in the 
highly non-linear regime. The lower dotted curve shows the linear power 
$\Delta_L^2(k)/k$. 
{\it Right panel:} The variance $\lag\Map^2\rag$ of the aperture-mass as 
a function of smoothing angle $\theta_s$ for the SNAP survey. We again 
display the cases $f_2=0.85$ (lower dashed curve), $f_2=1$ (middle solid line) 
and $f_2=1.15$ (upper dotted curve).}
\label{FigXi}
\end{figure}

For illustration purposes, we show in Fig.~\ref{FigXi} the non-linear power 
$\Delta^2(k)/k$ (left panel, as a function of comoving wavenumber $k$) at 
redshift $z=1$ and the variance $\lag\Map^2\rag$ (right panel, as a function
of smoothing angle $\theta_s$) for the SNAP survey described in 
section~\ref{Survey} below with no redshift binning. We display
the curves obtained for the cases $f_2=0.85,1,1.15$ from bottom to top.
This corresponds to a variation of about $\pm 10\%$ for the power-spectrum 
$P(k)$ in the non-linear regime. Thus, the range $f_2=1\pm 0.15$ describes
roughly the current uncertainty on the non-linear power-spectrum in the
range of interest. Note that $f_2$ is not merely
a multiplicative factor to the non-linear power-spectrum as the recipe from
Peacock \& Dodds (1996) also involves a rescaling of length scales.
Here we defined the 3-D power per logarithmic
interval $\Delta^2(k)=4\pi k^3 P(k,z)$. Because of the integration along
the line of sight the power relevant for weak lensing observables is actually
$\Delta^2(k)/k$, and the typical wavenumber associated with the angle 
$\theta_s$ is $k \sim 4/(\De\theta_s)$, see Munshi et al. (2004).
The lower dotted curve in left panel of Fig.~\ref{FigXi} shows the linear
power $\Delta_L^2(k)/k$. We can check that at low wavenumbers and at large 
angular scales which probe the linear regime the dependence on $f_2$ 
disappears.

\begin{figure}
\begin{center}
\epsfxsize=6 cm \epsfysize=5 cm {\epsfbox{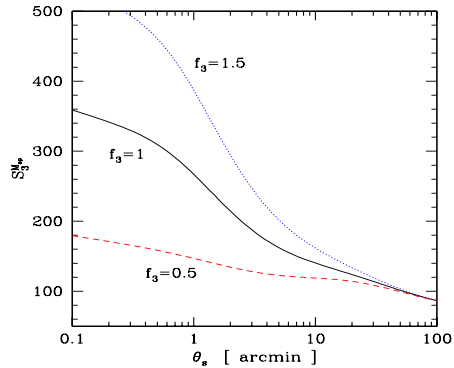}} 
\end{center}
\caption{The skewness $S_3^{\Map}$ of the aperture-mass as as a function of
the smoothing angle $\theta_s$ for the SNAP survey. We display the cases 
$f_3=0.5$ (lower dashed curve), $f_3=1$ (middle solid line) and $f_3=1.5$ 
(upper dotted curve) which correspond to a $50\%$ variation of the
skewness in the highly non-linear regime.}
\label{FigS3}
\end{figure}

For higher-order correlations of the matter density field we use the stellar
model (\ref{stellar}) described in Valageas et al. (2004a). This is
actually identical to the minimal tree-model up to third-order moments
(Munshi et al. 2004). In particular, for the skewness $S_3$ of the 3-D matter
density field we interpolate between the quasi-linear limit $S_3^{\rm QL}$ and
the non-linear prediction $S_3^{\rm HEPT}$ of HEPT (Scoccimarro et al. 1998).
In order to investigate the sensitivity to the non-linear fit used for $S_3$
we introduce a parameter $f_3$ and we define the skewness $S_3^{\rm NL}$ 
reached in the highly non-linear regime as:
\beq
S_3^{\rm NL} = f_3 S_3^{\rm HEPT} .
\label{f3}
\eeq
Then, on linear scale we use $S_3=S_3^{\rm QL}$ while on highly non-linear 
scales we use $S_3=S_3^{\rm NL}$, see Munshi et al. (2004) for details.

We show in Fig.~\ref{FigS3} the skewness 
$S_3^{\Map}=\lag\Map^3\rag/\lag\Map^2\rag^2$ as a function of
the smoothing angle $\theta_s$ for the cases $f_3=0.5,1,1.5$, for the SNAP 
survey described in section~\ref{Survey} below with no redshift 
binning. Thus we now consider a variation of $\pm 50\%$ on $f_3$ and $S_3$
since the skewness is not known to better than $50\%$ in the highly non-linear
regime (the actual current uncertainty may actually be even larger).
Of course, since the aperture-mass is linear over the density fluctuations 
within our weak-lensing approximation (\ref{Mapdelta}) this yields a 
$50\%$ variation for $S_3^{\Map}$ at small angular scales. We can check again 
that at large angular scales which probe the quasi-linear regime the 
dependence on $f_3$ disappears.

Thus, the parameters $f_2$ and $f_3$ allow us to evaluate the sensitivity of
weak lensing results onto the amplitude of the two-point and three-point
matter correlations in the non-linear regime, where they are not known to 
up to high accuracy yet. Unless otherwise stated, we use $f_2=1$ and $f_3=1$
which correspond to the fits obtained from Peacock \& Dodds (1996) and from
Scoccimarro et al. (1998).

\subsection{Survey parameters}
\label{Survey}

Hereafter, we adopt the characteristics of the SNAP mission as given in 
Refregier et al.(2004). More precisely, we consider the ``Wide'' survey where 
the redshift distribution of galaxies is given by:
\beq
n(z_s) \propto z_s^2 \; e^{-(z_s/z_0)^2} \;\;\; \mbox{and} \;\;\; 
z_0=1.13, \;\;\; z_{\rm max}=3 .
\label{nzSNAP}
\eeq
The variance in shear due to intrinsic ellipticities and measurement errors is 
$\sigma_*=\lag|\epsilon_*|^2\rag^{1/2}=0.31$. The survey covers an area 
$A=300$ deg$^2$ and the surface density of usable galaxies is 
$n_g=100$ arcmin$^{-2}$. In order to extract some information from the 
redshift dependence of weak lensing effects we also divide the ``Wide'' SNAP 
survey into two redshift bins: $0<z_s<z_*$ and $z_*<z_s<z_{\rm max}$.
We choose $z_*=1.23$, which corresponds roughly to the separation provided
 by the SNAP filters and which splits the ``Wide'' SNAP survey into two 
samples with the same number of galaxies (hence $n_g=50$ arcmin$^{-2}$). 
Note that one cannot use too many redshift bins at it decreases the number 
of source galaxies associated with each subsample (for the aperture mass we 
could still obtain good results with three bins but we shall restrict 
ourselves to two redshift bins in this paper). The redshift bins that we use 
are similar to those which were used by Refregier et al.(2003) using 
photometric redshifts, except that they have a sharp cutoff and 
non-overlapping source distributions. Note that using overlapping source 
distributions (over redshift) would increase the cross-correlations.

\section{Dependence of low-order estimators on Cosmology}
\label{Dependence-of-low-order-estimators}

\begin{figure}
\begin{center}
\epsfxsize=4.15 cm \epsfysize=4.5 cm {\epsfbox{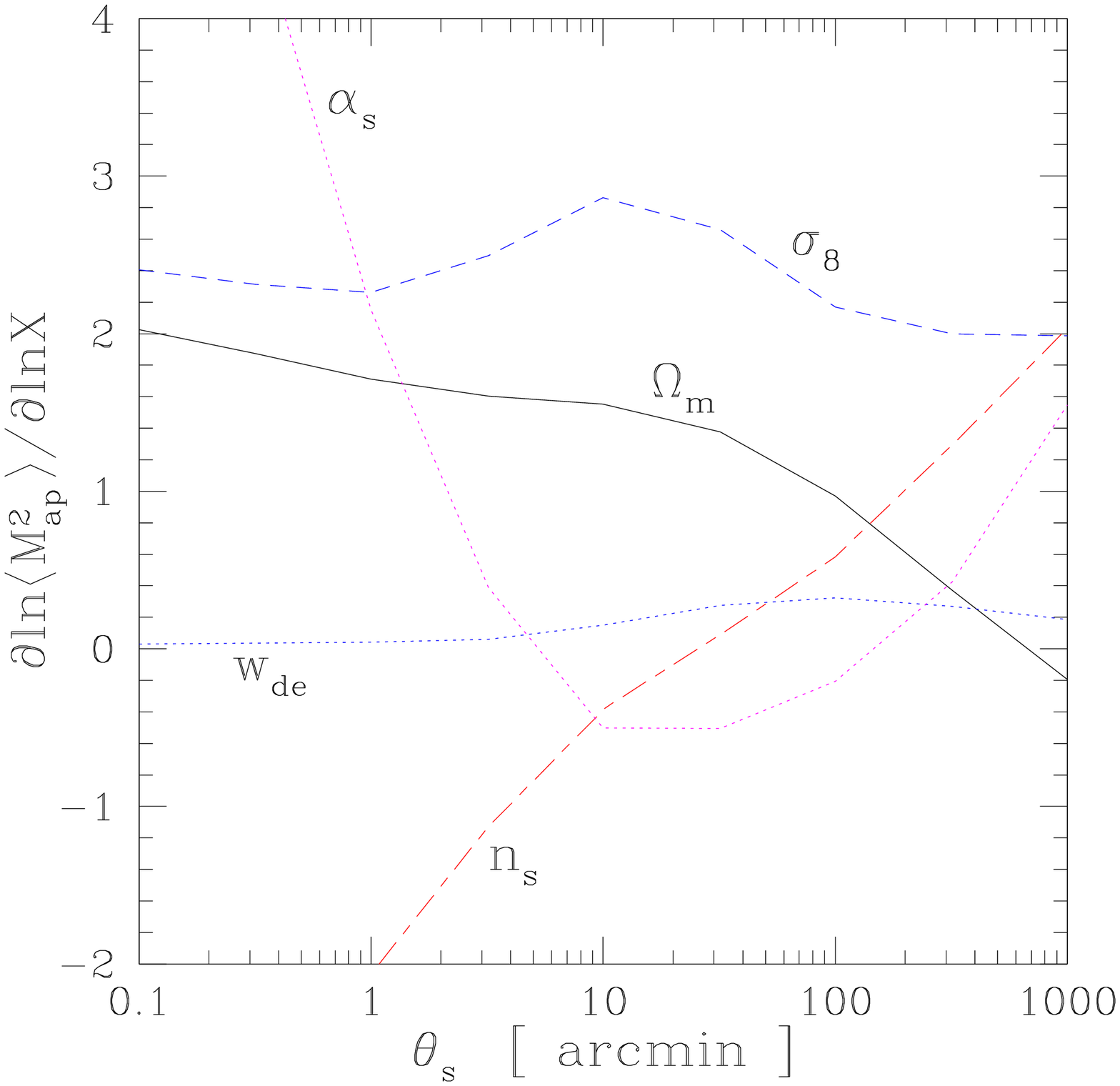}} 
\epsfxsize=4.15 cm \epsfysize=4.5 cm {\epsfbox{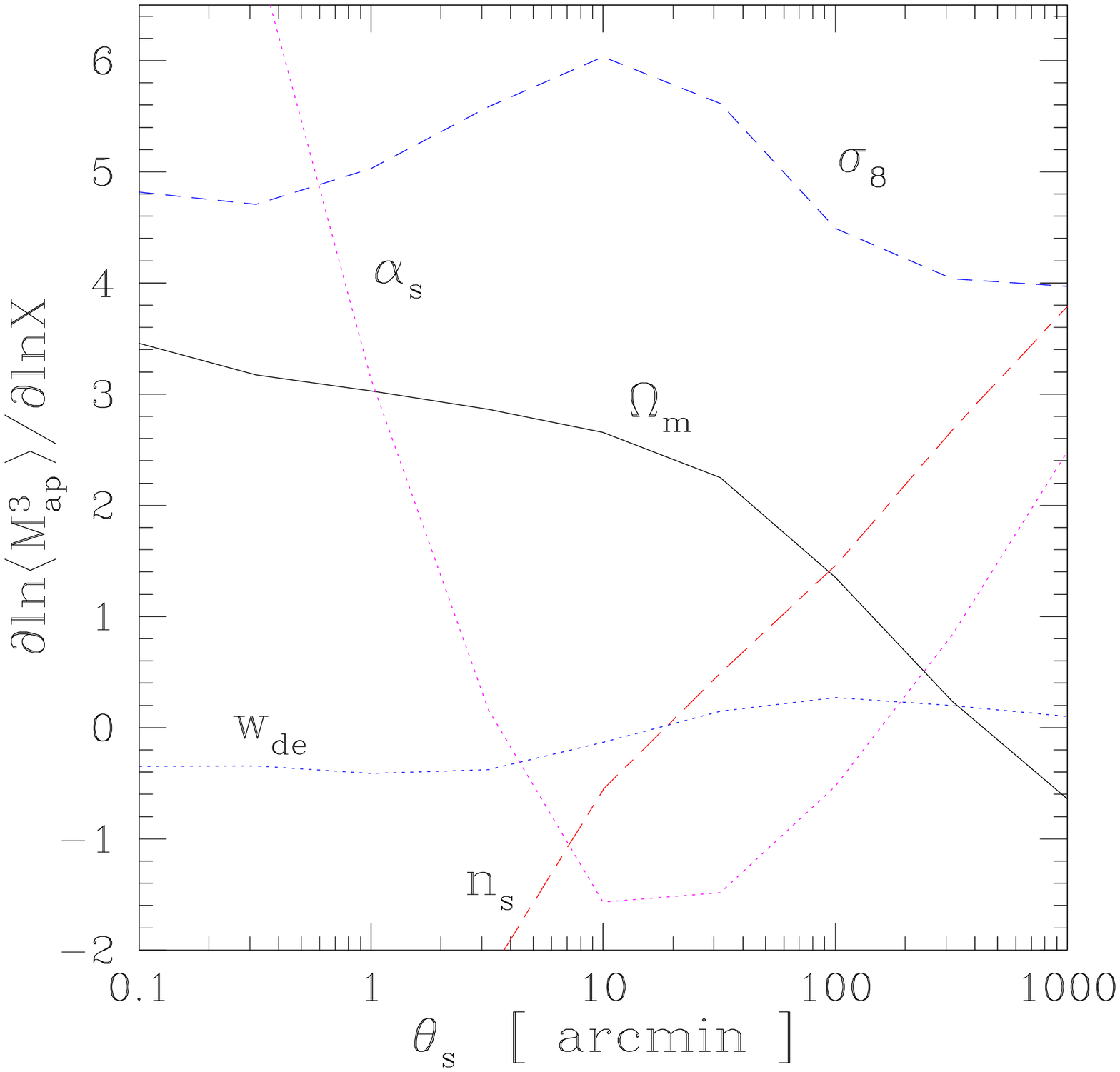}}\\
\epsfxsize=4.15 cm \epsfysize=4.5 cm {\epsfbox{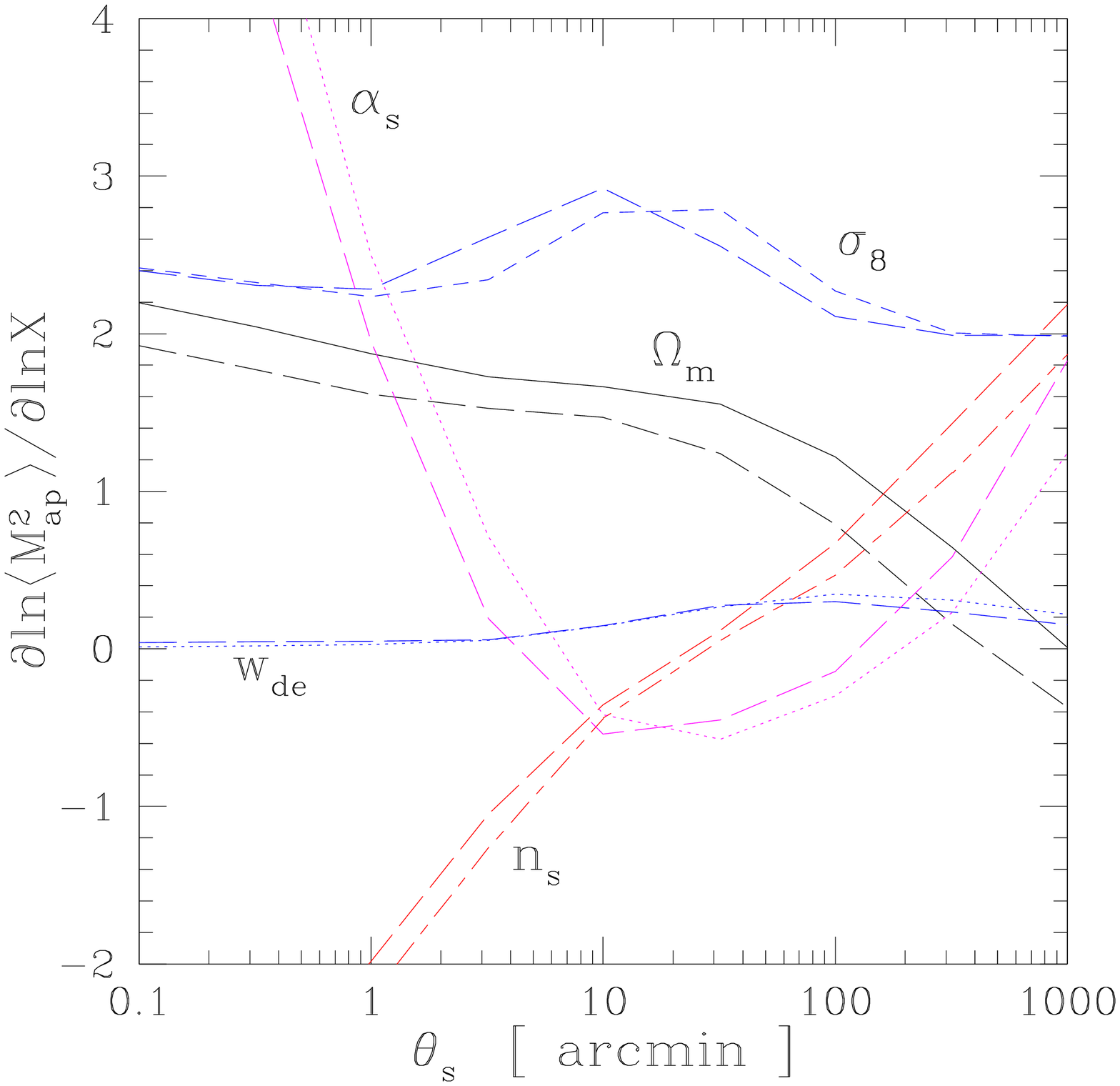}} 
\epsfxsize=4.15 cm \epsfysize=4.5 cm {\epsfbox{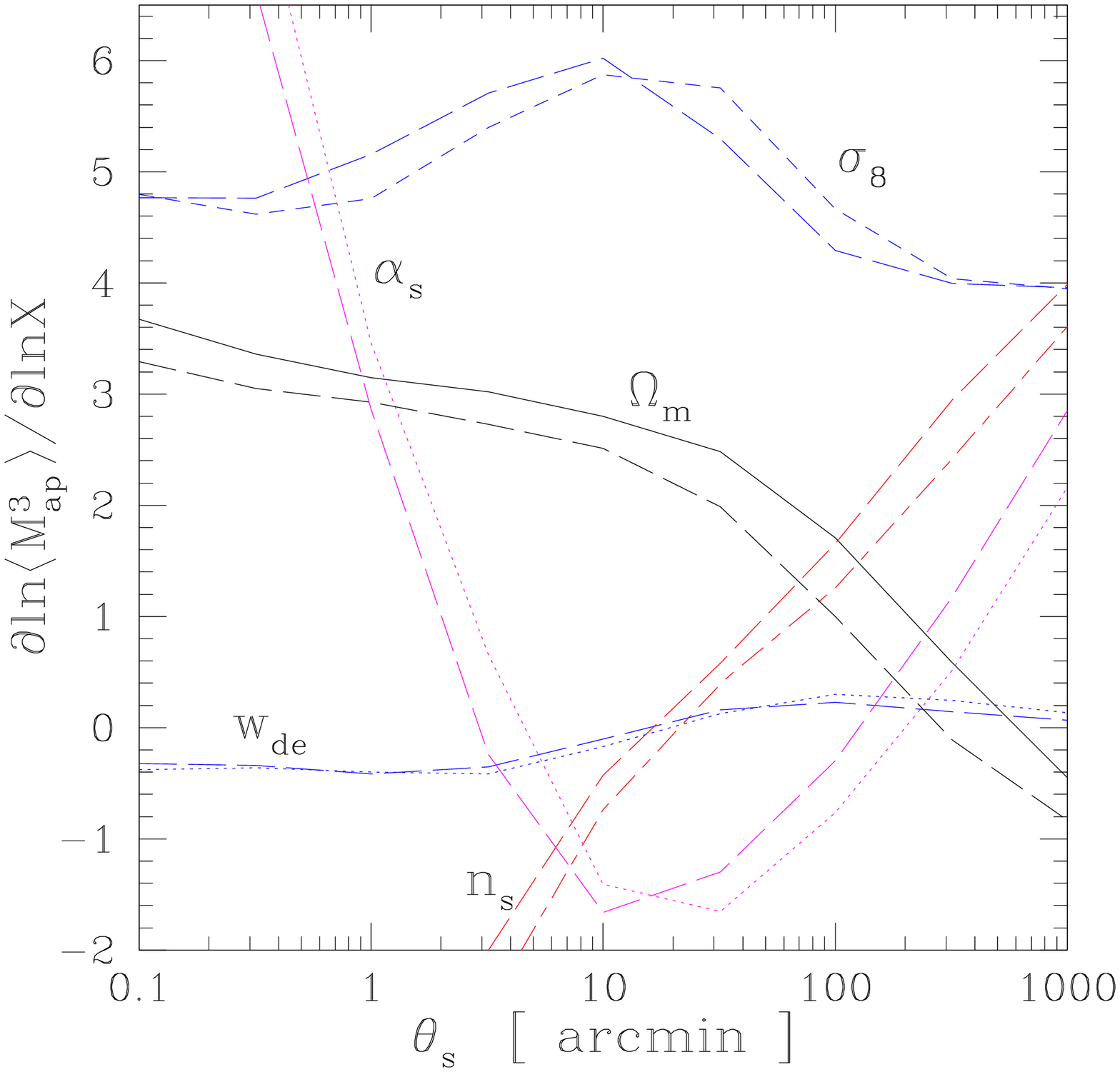}}
\end{center}
\caption{The logarithmic derivative 
$\partial\ln\lag\Map^p\rag_c/\partial\ln X$ 
of low order cumulants with respect to cosmological parameters $X$. 
{\it Upper left panel:} The dependence of the variance $\lag\Map^2\rag$ on
$\Om$ (solid line), $\sigma_8$ (dashed line), $\wde$ (dotted line),
$n_s$ (dot dashed line) and $\alpha_s$ (dotted line), as a function of 
angular scale $\theta_s$ with no redshift binning. For $\alpha_s$ we plot
$\partial\ln\lag\Map^2\rag_c/\partial\alpha_s$ since $\alpha_s=0$ in our 
fiducial model.
{\it Upper right panel:} Same as upper left panel for the third-order
cumulant $\lag\Map^3\rag_c$.
{\it Lower left panel:} Same as upper left panel but for two redshift bins.
The low redshift bin is shown by the same line styles as in upper left panel
whereas the high redshift bin is shown by the long dashed line.
{\it Lower right panel:} Same as lower left panel for the third-order
cumulant $\lag\Map^3\rag_c$.}
\label{logH}
\end{figure}

We show in Fig.~\ref{logH} the logarithmic derivative 
$\partial\ln\lag\Map^p\rag_c/\partial\ln X$ of cumulants of  order $p=2$ (left 
panels) and $p=3$ (right panels) with respect to cosmological parameters $X$,
with no redshift binning (upper panels) and with two redshift bins (lower 
panels). For the cosmological parameter $\alpha_s$ we merely plot
$\partial\ln\lag\Map^p\rag_c/\partial\alpha_s$ since $\alpha_s=0$ in our 
fiducial model. 
The rather large values obtained for these derivatives shows that weak
lensing effects could potentially constrain cosmological parameters up to
a good accuracy. By contrast, the small derivative with respect to $\wde$
shows that the equation of state of the dark energy component cannot be
measured up to a similar accuracy. As expected, since weak lensing effects
are proportional to matter density fluctuations, eq.(\ref{Mapdelta}), the
derivatives with respect to $\Om$ and $\sigma_8$ are positive over most
of the angular range $0.1'-1000'$. In particular, we can check that
$\lag\Map^2\rag_c \sim \sigma_8^2$ as expected from eq.(\ref{Mapdelta}).
On the other hand, we can check that $\partial\lag\Map^2\rag_c/\partial n_s>0$ 
at small scales which probe high wavenumbers $k$ and it crosses zero at about
$20'$ which corresponds to the normalisation of the power-spectrum.
Since these derivatives show different variations with the angular
scale $\theta_s$ for different cosmological parameters, one should be able 
to constrain simultaneously these parameters by using several angular
scales. Note that by looking over such a large range of angular scales
one can discriminate the behaviours of different cosmological parameters
which helps to remove degeneracies.
On the other hand, one can use the additional information provided by
higher-order cumulants, such as the third-order cumulant (whence the skewness)
shown in the right panels. However, we can see that for most parameters
$\lag\Map^3\rag_c$ behaves roughly as $\lag\Map^2\rag^2$, except for $\Om$
where there is a residual dependence which can be used to measure for
instance both $\Om$ and $\sigma_8$ (see also Bernardeau et al. 1997, as well
as Kilbinger \& Schneider 2005).
Alternatively, one can split the survey into two redshift bins and take
advantage of the different dependence on cosmology of weak lensing effects
associated with each bin. This is shown in both lower panels. Although the
curves are quite similar we shall check in sect.~\ref{Estimation} that
using such a redshift binning does indeed improve the constraints on
cosmological parameters.

\section{Covariance matrices of low-order estimators}
\label{Covnum}

In order to use the estimators $\cH_p$ to measure cosmological parameters,
through their dependence on cosmology displayed in Fig.~\ref{logH}, we need
the covariance matrices $C_{ij}$ introduced in sect.~\ref{Covariance}.
The latter are necessary to obtain the relevant error bars through a
Fisher matrix analysis (sect.~\ref{Formalism}) or a $\chi^2$ likelihood 
function (sect.~\ref{Fisherchi2}).

\subsection{Impact of noise and non-Gaussianities}
\label{Covnoise}

\begin{figure} 
\epsfxsize=8.1 cm \epsfysize=6 cm {\epsfbox{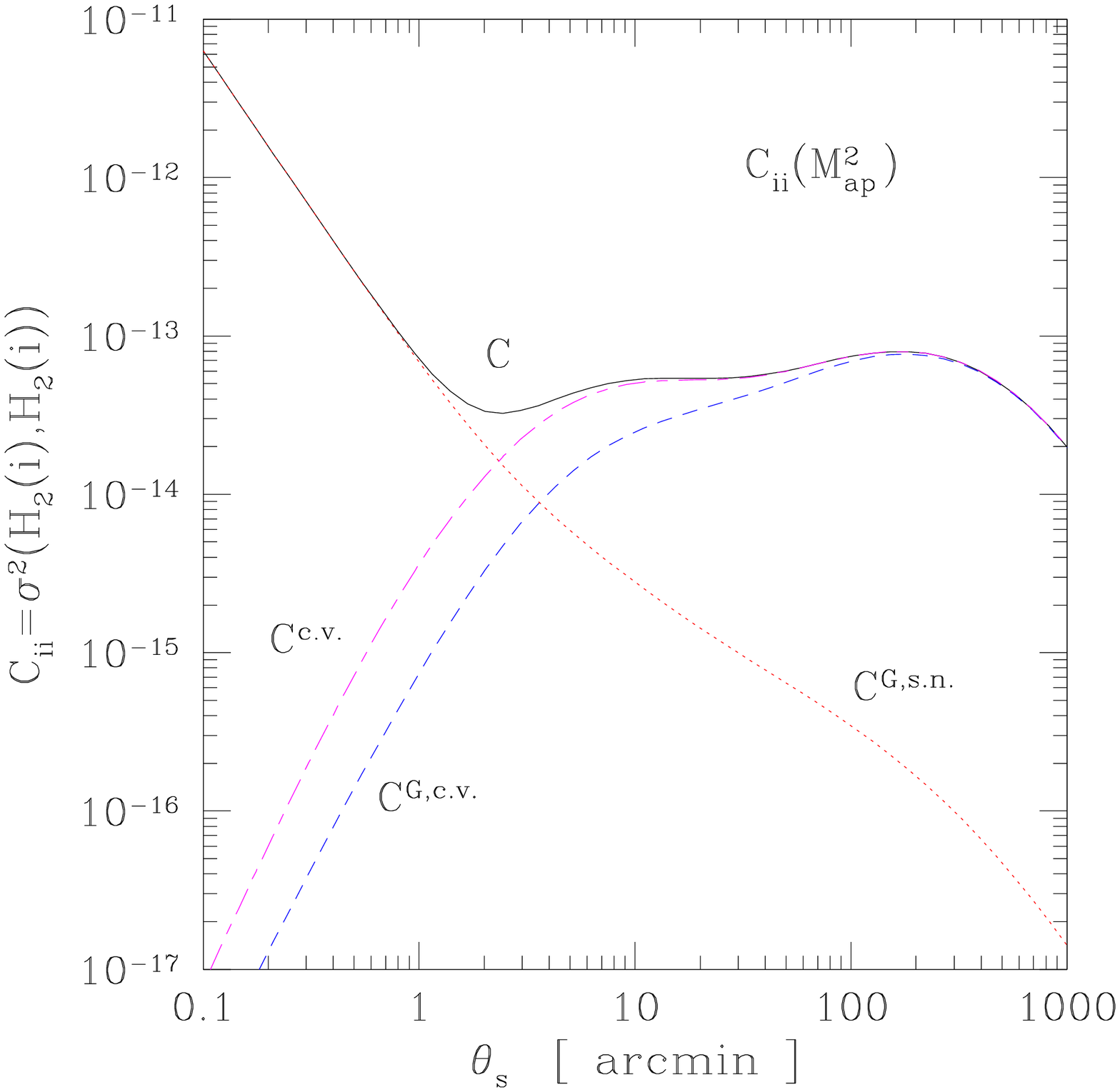}}
\caption{Covariance of the estimator $\cH_2(\theta_s)$ of the variance 
$\lag\Map^2\rag$, eqs.(\ref{cHp})-(\ref{Cij}). The solid line shows the full 
covariance $C_{ii}=\sigma^2(\cH_2(\theta_{si}),\cH_2(\theta_{si}))$ as a 
function of smoothing angular scale $\theta_{si}$, with no redshift binning. 
The dotted line displays the contribution 
$C^{G,s.n.}$ from the galaxy intrinsic ellipticity dispersion to the Gaussian 
part of $C$. The dashed line $C^{G,c.v.}$ is the cosmic variance contribution
to the Gaussian part of $C$ while $C^{c.v.}$ is the cosmic variance 
contribution to the full matrix $C$.}
\label{CovH2}  
\end{figure}

\begin{figure} 
\epsfxsize=8.1 cm \epsfysize=6 cm {\epsfbox{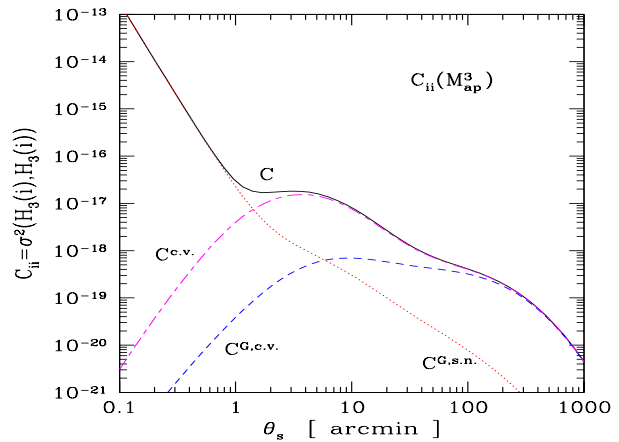}}
\caption{Covariance of the estimator $\cH_3(\theta_s)$ of the third-order 
cumulant $\lag\Map^3\rag_c$, eq.(\ref{cHp}). The various line styles 
show different contributions to the full covariance
$C_{ii}=\sigma^2(\cH_3(\theta_{si}),\cH_3(\theta_{si}))$ (solid line)
as in Fig.~\ref{CovH2}.}
\label{CovH3}  
\end{figure}

\begin{figure}
\begin{center}
\epsfxsize=4.15 cm \epsfysize=4.5 cm {\epsfbox{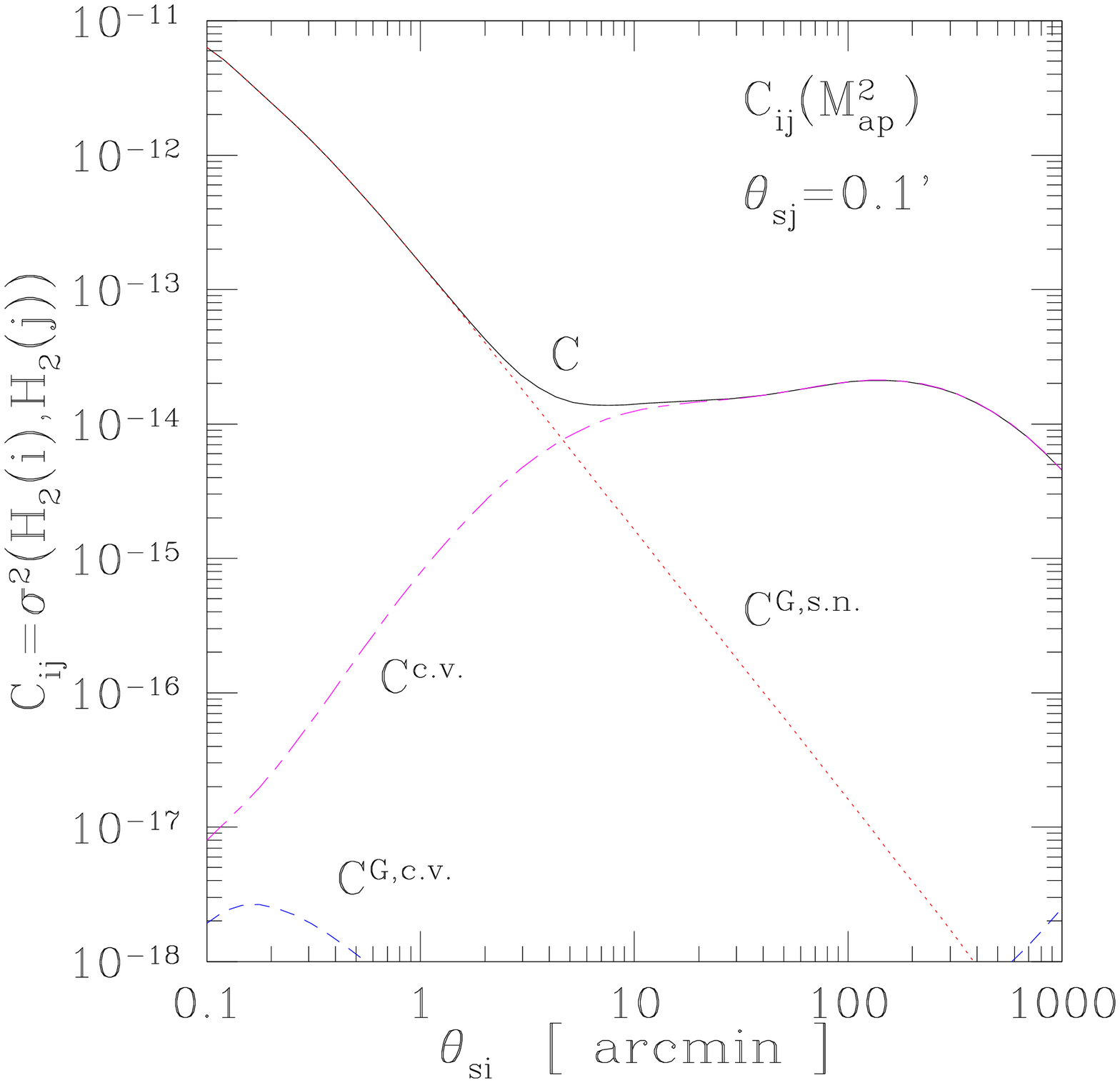}} 
\epsfxsize=4.15 cm \epsfysize=4.5 cm {\epsfbox{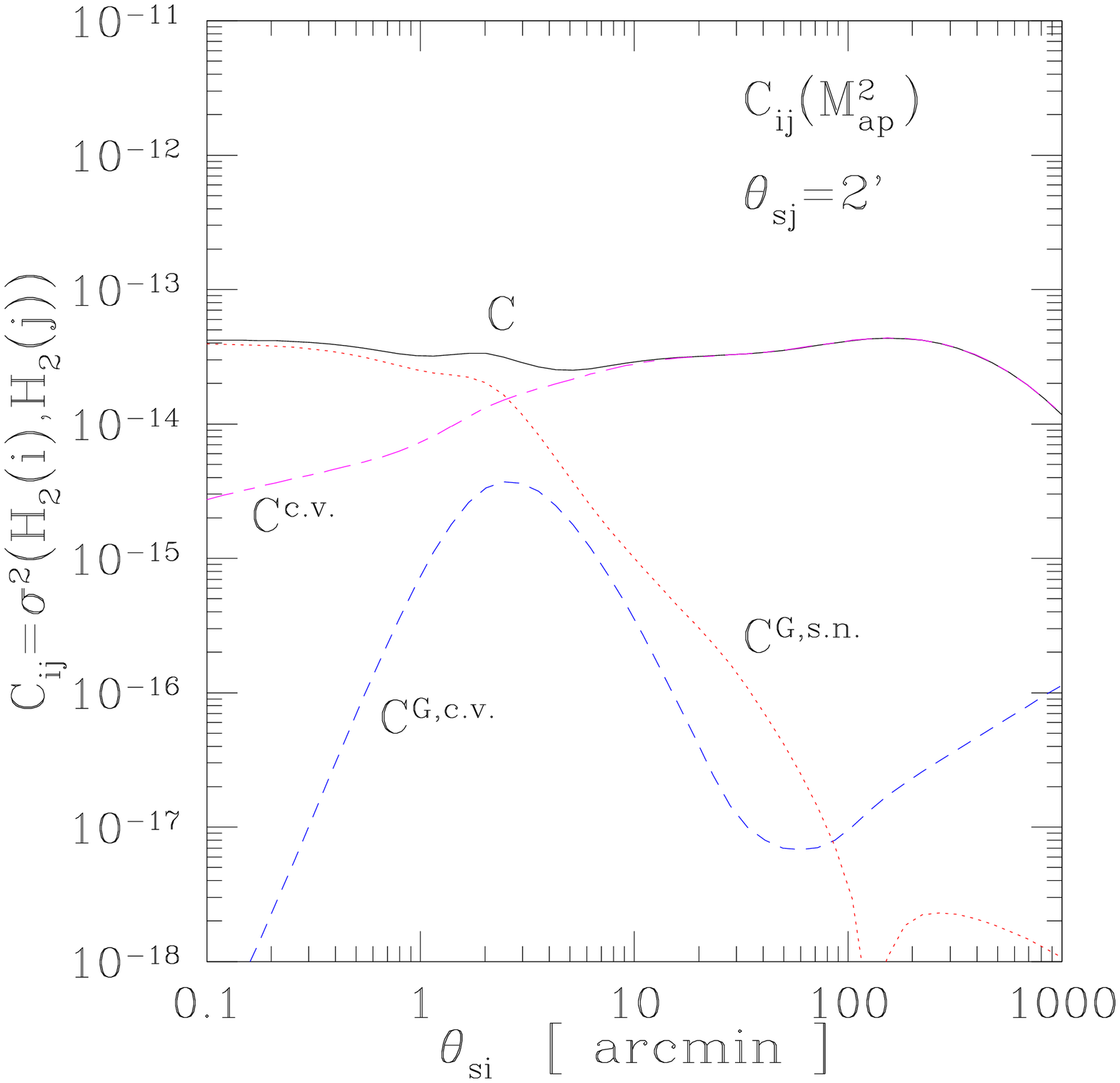}}\\
\epsfxsize=4.15 cm \epsfysize=4.5 cm {\epsfbox{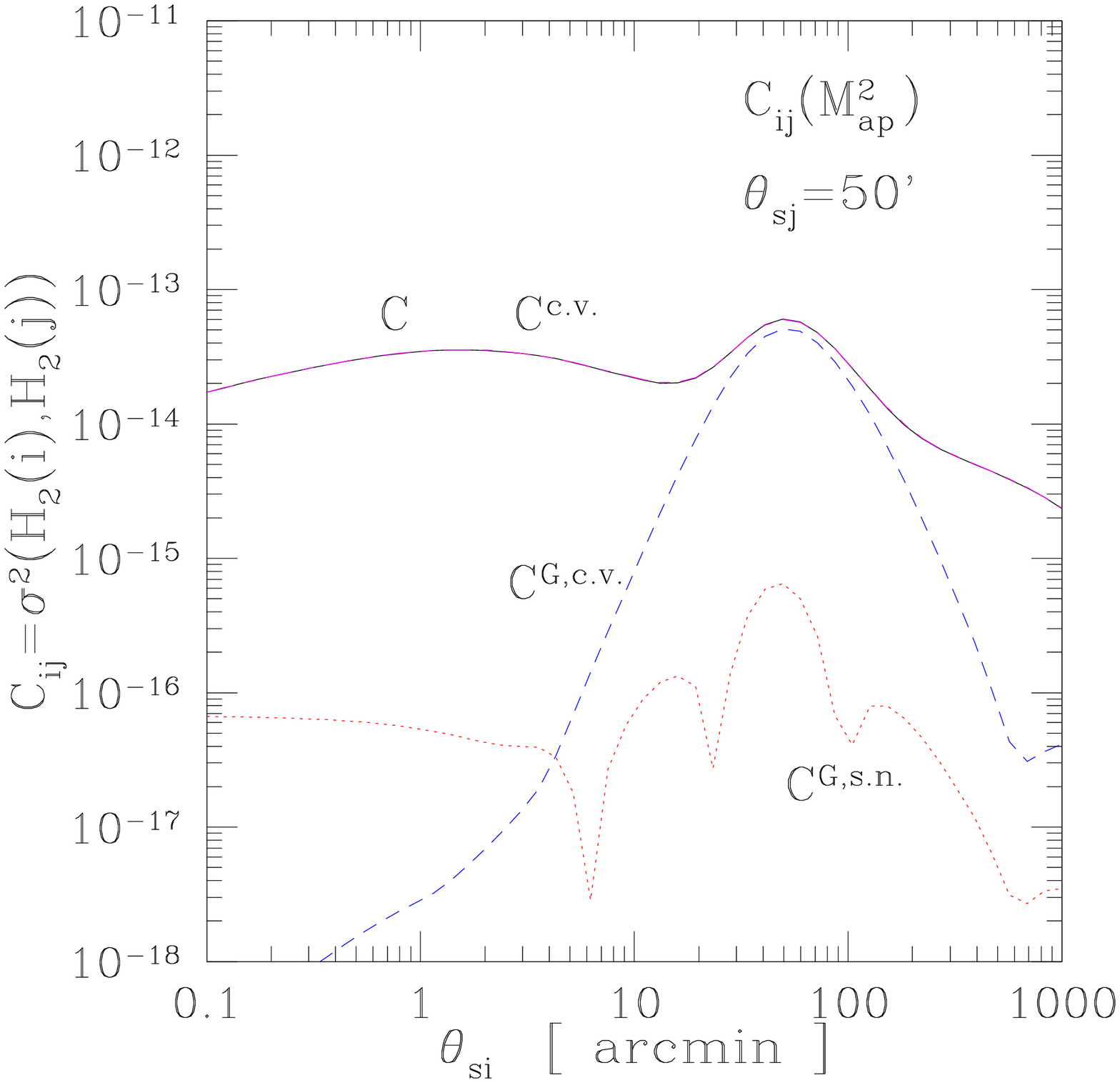}} 
\epsfxsize=4.15 cm \epsfysize=4.5 cm {\epsfbox{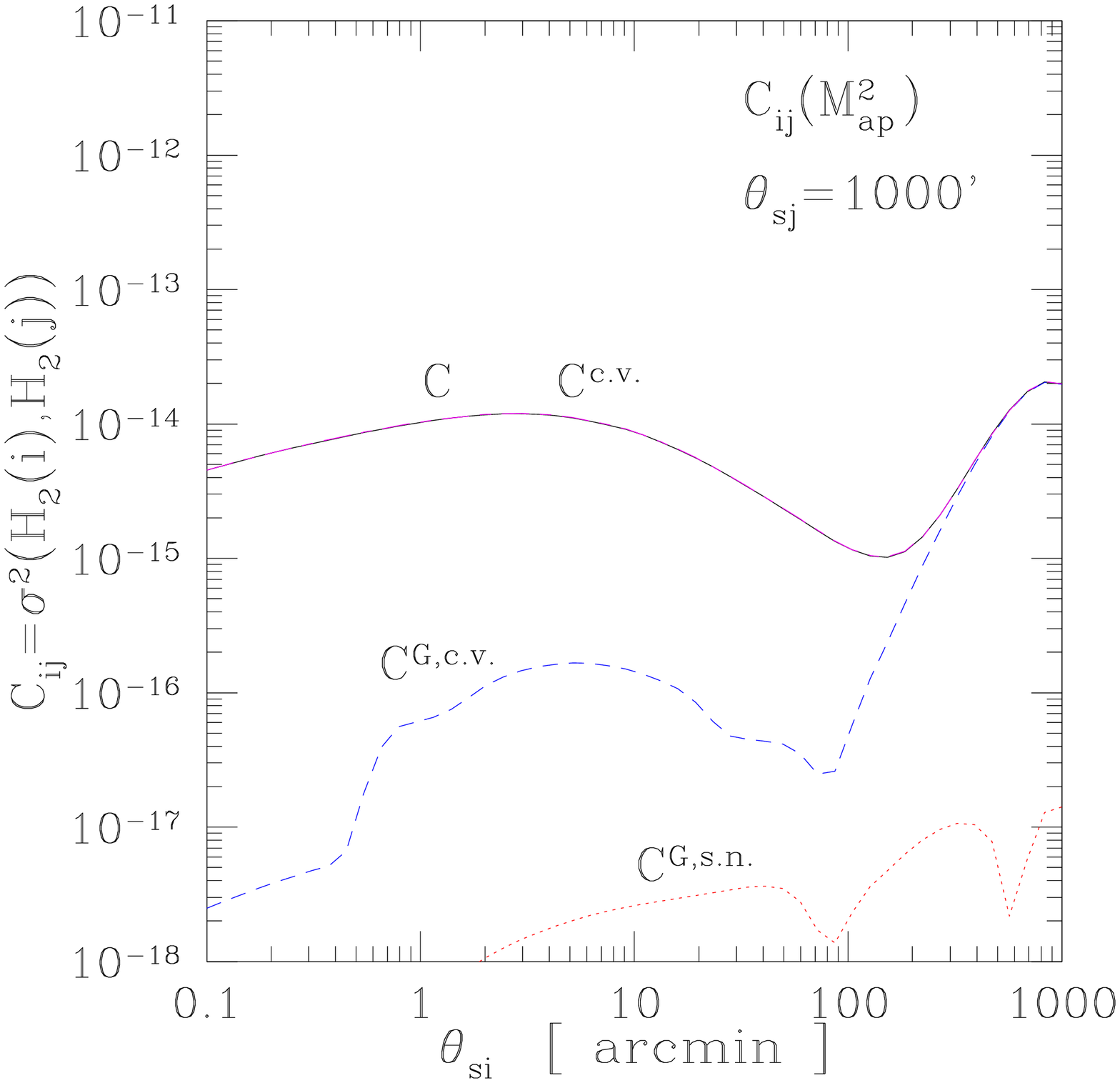}}
\end{center}
\caption{Covariance $C_{ij}$ of the estimator $H_2(\theta_s)$ of the variance 
$\lag\Map^2\rag$, eqs.(\ref{cHp})-(\ref{Cij}), for different angular
scales as a function of $\theta_{si}$ at fixed $\theta_{sj}$, with no
redshift binning. The various line 
styles show different contributions to the full covariance as in 
Fig.~\ref{CovH2}. The four panels correspond to $\theta_{sj}=0.1'$ (upper left
panel), $\theta_{sj}=2'$ (upper right panel), $\theta_{sj}=50'$ (lower left
panel) and $\theta_{sj}=1000'$ (lower right panel).}
\label{CovHij}
\end{figure}

We first show in Fig.~\ref{CovH2} the covariance matrix $C_{ij}$ of the
estimator $\cH_2(\theta_s)$ of the variance $\lag\Map^2\rag$, see 
eqs.(\ref{cHp})-(\ref{Cij}), along the diagonal $i=j$ with no redshift binning
($\theta_{si}=\theta_{sj}$). The full solid line shows the full covariance
$C$ while other line styles correspond to the specific contributions
$C^{G,s.n.}$, $C^{G,c.v.}$ and $C^{c.v.}$, defined in 
eqs.(\ref{sigma2Gcv})-(\ref{sigma2cv}). As expected, we check that at small
angular scales $\theta_s<1'$ the covariance $C$ is dominated by the shot-noise
due to the galaxy intrinsic ellipticity dispersion, whereas at large scales
$\theta_s>2'$ the covariance is dominated by the cosmic variance (i.e. the
error bar due to the finite size of the survey). We can also check that
the shot noise contribution $C^{G,s.n.}$ grows as $1/\theta_s^2$ at smaller
scales, which is proportional to the number of distinct patches of radius
$\theta_s$ in the survey. The slow rise of the cosmic variance term $C^{c.v.}$
yields a broad plateau for the full covariance, from $1'$ up to $1000'$, while
non-Gaussianities become negligible above $30'$. We can note that even at 
smallest scales non-Gaussian terms are only twice larger than Gaussian terms
so that the Gaussian part $C^G=C^{G,s.n.}+C^{G,c.v.}$ of the covariance
matrix is always a good approximation along the diagonal 
($\theta_{si}=\theta_{sj}$) for two-point estimators such as $\cH_2$.

Next, we display in Fig.~\ref{CovH3} the covariance matrix $C_{ii}$ of
the estimator $\cH_3(\theta_s)$ of the third-order cumulant $\lag\Map^3\rag_c$,
again along the diagonal $i=j$ with no redshift binning 
($\theta_{si}=\theta_{sj}$), see eq.(\ref{sigma33}). We find again that the 
shot-noise due to the galaxy intrinsic ellipticities dominates below $2'$.
However, it now grows as $1/\theta_s^4$ rather than $1/\theta_s^2$ at smallest
scales (this is due to the term $Q_{ij}^3$ in eq.(\ref{sigma33})). 
Above $1'$ the covariance is dominated by the cosmic variance but
in contrast with Fig.~\ref{CovH2} we now find that non-Gaussian terms are
larger than Gaussian terms by an order of magnitude around $\theta_s \sim 3'$
(for the cosmic variance part).
This is not really surprising since $\cH_3$ itself is an estimator of 
non-Gaussianities. However, above $60'$ non-Gaussiannities become negligible
again as we probe the quasi-linear regime.

We show in Fig.~\ref{CovHij} the covariance $C_{ij}$ of the estimator 
$\cH_2(\theta_s)$ of the variance $\lag\Map^2\rag$ for different angular
scales $\{\theta_{si},\theta_{sj}\}$ with no redshift binning (i.e. we probe 
$C_{ij}$ out of the diagonal whereas Fig.~\ref{CovH2} was restricted to the 
diagonal $i=j$). 
The line styles are as in Fig.~\ref{CovH2}. The wiggles in the lower panels
for $C^{G,s.n.}$ correspond to a change of sign (hence we actually plot 
$|C^{G,s.n.}|$) because at large and different angular scales the first term
in eq.(\ref{sigma2Gsn}) can make $C^{G,s.n.}$ negative.When both angular 
scales 
are small the covariance is dominated by the intrinsic ellipticity noise
(both upper panels with $\theta_{si}<2'$) but as soon as one scale is larger
than $10'$ the covariance is dominated by the cosmic variance (both upper 
panels with $\theta_{si}>2'$ and both lower panels for all $\theta_{si}$).
Then, except at large scales where $\theta_{si} \sim \theta_{sj}$ the
cosmic variance is dominated by the non-Gaussian terms which can be several
orders of magnitude larger than the Gaussian contribution. Note that this
is quite different from Fig.~\ref{CovH2} which showed that non-Gaussian
contributions were not very important along the diagonal. Moreover,
the non-Gaussian contribution to $C_{ij}$ shows a broad plateau as we
go farther from the diagonal. This feature, 
which makes the covariance matrix very broad about the diagonal and
reflects a strong correlation between various angular scales, leads to
difficulties for the estimation of cosmological parameters since the
covariance matrix $C_{ij}$ cannot be easily inverted. 

\begin{figure*}
\protect\centerline{
\epsfysize = 3.75truein
\epsfbox[0 0 383 395]
{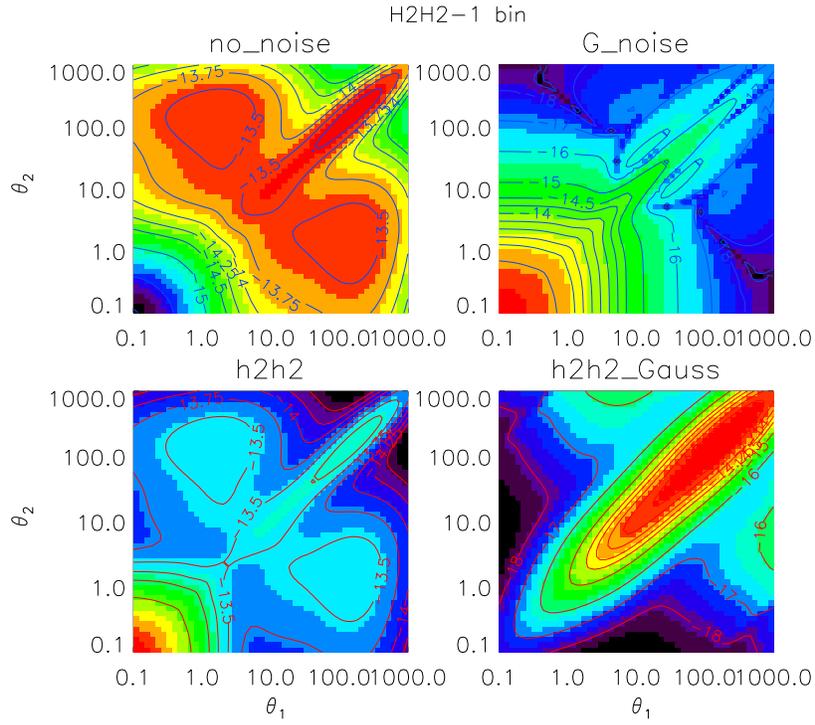}}
\caption{Covariance matrix for the estimator $\cH_2$ for different angular 
scales. We plot contours of constant $\log C$.
{\it Lower left panel:} the full covariance $C$.
{\it Upper left panel:} the cosmic variance contribution $C^{c.v.}$ (i.e.
the shot noise due to the galaxy intrinsic ellipticity dispersion is set to 
zero).
{\it Lower right panel:} the Gaussian contribution $C^{G,c.v.}$ to the
cosmic variance.
{\it Upper right panel:} the shot noise contribution $C^{G,s.n.}$ to the
Gaussian part of $C$.}
\label{fig:cov_h2h2}
\end{figure*}

\begin{figure*}
\protect\centerline{
\epsfysize = 3.75truein
\epsfbox[0 0 383 395]
{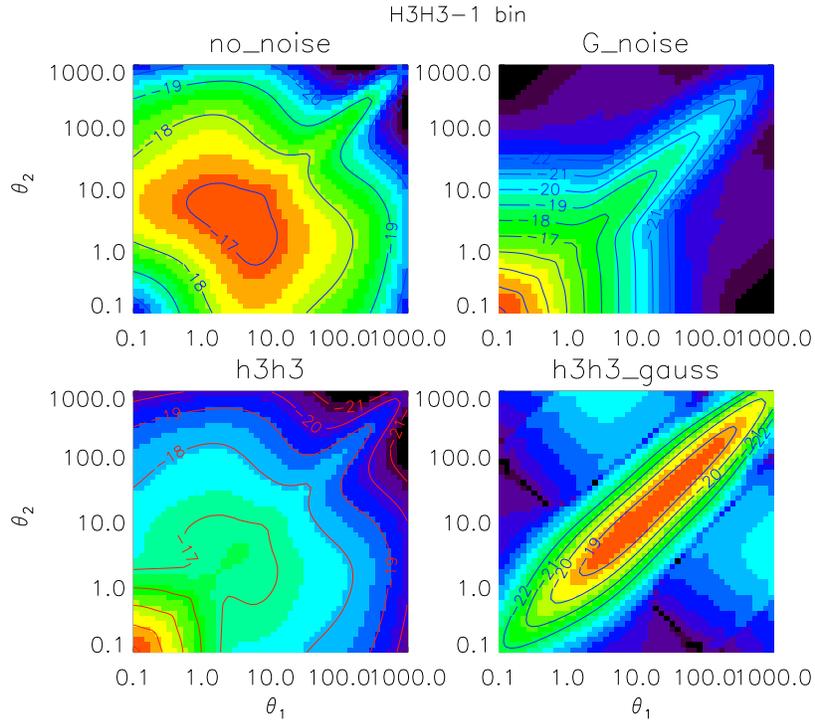}}
\caption{Same as previous figure but for for the estimator $\cH_3$. 
See text for discussion.}
\label{fig:cov_h3h3}
\end{figure*}

The full covariance matrix along with its various components are shown 
in Fig.~\ref{fig:cov_h2h2} for $\cH_2$.
In agreement with Figs.~\ref{CovH2}-\ref{CovHij} the upper right panel which
displays $C^{G,s.n.}$ shows that the shot noise increases fastly at
smaller angular scales. Its contribution to the covariance is restricted to
the diagonal $\theta_{s1}=\theta_{s2}$ above $10'$ while it mostly depends
on $\max(\theta_{s1},\theta_{s2})$ when both scales are below $10'$.
On the other hand, the Gaussian contribution $C^{G,c.v.}$ to the cosmic 
variance (lower right panel) is always restricted close to the diagonal and
increases at large scales. The upper left panel which displays $C^{c.v.}$
clearly shows that non-Gaussianities bring a significant broadening 
to the covariance matrix which is no longer diagonal-dominated. Finally,
the lower left panel which displays the full covariance matrix $C$ reflects
these various factors and shows a very broad shape with a steep increase at
small angular scales due to the shot-noise. Fig.~\ref{fig:cov_h3h3} which
shows the covariance matrix for $\cH_3$ exhibits a similar behaviour albeit
with a stronger impact of non-Gaussianities.

\subsection{Signal to noise ratios}
\label{Signal-to-noise-ratios}

\begin{figure} 
\epsfxsize=8.1 cm \epsfysize=6 cm {\epsfbox{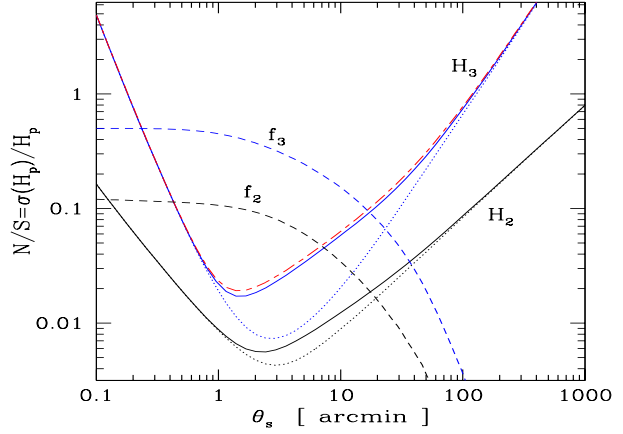}}
\caption{Inverse of the signal to noise ratios of the estimators $\cH_2$ and
$\cH_3$ of the second-order and third-order cumulants of the aperture-mass.
The solid lines show the ratios $\sigma(H_p(i),H_p(i))/H_p(i)$ as a function
of the angular scale $\theta_{si}$ for $p=2$ (curve labeled $H_2$) and 
$p=3$ (curve labelled $H_3$) with no redshift binning. The covariance 
$\sigma^2=C_{ii}$ also corresponds to the solid line in 
Figs.~\ref{CovH2}-\ref{CovH3}. The dotted lines show the ratios obtained when 
we only include Gaussian terms in the covariance $C_{ii}$ while the dashed 
lines correspond to the uncertainty associated with a $15\%$ variation of 
$f_2$ with respect to $\cH_2$ and a $50\%$ variation of $f_3$ with respect to 
$\cH_3$. The dot-dashed line which follows closely the curve obtained for 
$\cH_3$ shows the noise/signal ratio when we use the estimator $\cM_3$ 
(eq.(\ref{cMp})) instead of $\cH_3$.}
\label{NSH23}  
\end{figure}

Next, we plot in Fig.~\ref{NSH23} the inverse $N/S$ of the signal to noise 
ratios of the estimators $\cH_2$ and $\cH_3$ of the second-order and 
third-order cumulants of the aperture-mass. We show the ratios 
$\sigma(H_p(i),H_p(i))/H_p(i)$ without redshift binning (solid lines). They
grow at small scales because of the shot noise due to the galaxy intrinsic
ellipticity dispersion and at large scales because of cosmic variance.
We also display the noise to signal ratios $N/S$ obtained when we only
include the Gaussian terms $C^G$ in the covariance $C_{ii}$ (dotted lines). 
This corresponds to $C^G=C^{G,s.n.}+C^{G,c.v.}$, that is the sum of dotted 
lines 
and dashed lines shown in Figs.~\ref{CovH2}-\ref{CovH3}. In agreement with
sect.~\ref{Covnoise} we find that non-Gaussian terms only make a difference
in the intermediate range $1'-100'$ and that this effect is quite modest
for the estimator $\cH_2$ of the variance $\lag\Map^2\rag$. For $\cH_3$
the noise ratio can be increased by a factor $3$ around $3'$. As discussed
in sect.~\ref{Covnoise} non-Gaussian terms are mainly important for 
cross-correlations between different angular scales. 

It is interesting to note that the angular scale at which the highest 
signal/noise ratio is achieved shifts to smaller angular scales when we 
include the contribution to cosmic variance from non-Gaussianities.
Thus ignoring non-Gaussian terms in cosmic variance not only
gives a wrong impression of higher signal to noise ratio, it also gives a wrong
estimate of the angular scale where this is achieved. More detailed analysis
of how these plots change with variations in survey charecteristics are
presented in the appendix, see fig.~\ref{NSH23all} and fig.~\ref{NSH23ground}. 
The very high S/N achieved by surveys such as SNAP for $\cH_3$ clearly 
indicates the potential of weak lensing surveys in studying even higher order 
non-Gaussianity such as $\cH_4$ which estimates the kurtosis of underlying 
mass distribution. The dot-dashed line which follows closely the curve 
obtained for $\cH_3$ shows the noise/signal ratio when we use the estimator 
$\cM_3$ (eq.(\ref{cMp})) instead of $\cH_3$. Thus, in agreement with
Valageas et al.(2005) we find that for third-order estimators there is not
much gain to be obtained by switching from $\cM_3$ to $\cH_3$. However,
for higher-order non-Gaussianities like $\cH_4$ the improvement can be
quite significant (see Figs.4 and 5 of Valageas et al. 2005).

Finally, the dashed lines
in Fig.~\ref{NSH23} show the noise associated with the current uncertainty
on the non-linear regime of gravitational clustering, as described by the
parameters $f_2$ and $f_3$. Thus, the curve labeled $f_2$ shows 
$N/S=\Delta \cH_2/\cH_2$ with $\Delta\cH_2=(\cH_2(f_2=1.15)-\cH_2(f_2=0.85))/2$
associated with a $15\%$ variation of $f_2$ as described in Fig.~\ref{FigXi}.
In agreement with Fig.~\ref{FigXi} and Fig.~\ref{NSH23} this corresponds 
roughly
to a $10\%$ uncertainty on the non-linear power-spectrum. We can see that this
is actually the dominant source of noise over the range $0.2'-20'$. Next, the 
curve labelled $f_3$ shows $N/S=\Delta \cH_3/\cH_3$ with 
$\Delta\cH_3=(\cH_3(f_3=1.5)-\cH_3(f_3=0.5))/2$ associated with a $50\%$ 
variation of $f_3$ as described in Fig.~\ref{FigS3}.
In agreement with Fig.~\ref{FigS3} and Fig.~\ref{NSH23} this corresponds 
roughly
to a $50\%$ uncertainty on the non-linear skewness and third-order cumulant.
This is the main source of noise over the range $0.3'-20'$.
Thus, in agreement with some previous studies we find that current theoretical
uncertainties on non-linear gravitational clustering are the limiting
factor to constrain cosmological parameters from large weak-lensing surveys
with characteristics similar to the SNAP mission.

\subsection{Cross-correlation of redshift bins}
\label{Crossredshift}

\begin{figure*}
\protect\centerline{
\epsfysize = 2.25truein
\epsfbox[0 0 510 190]
{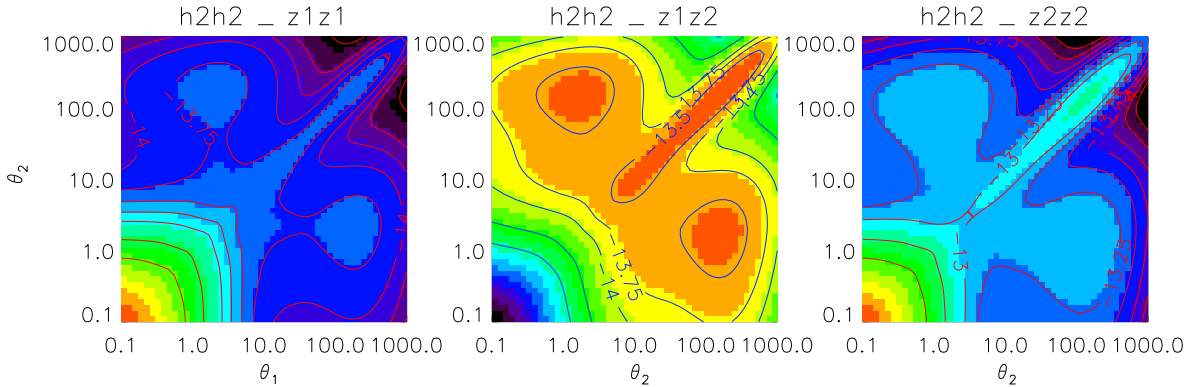}}
\caption{Covariance and cross-correlation is displayed for two different 
redshift bins.
Left panel shows the covariance of $\cH_2$ for lower redshift bin and
right panel corresponds to higher redshift bin. Middle panel panel shows the 
cross-correlation between both redshift bins.}
\label{fig:cov_h2h2_zbin}
\end{figure*}

\begin{figure*}
\protect\centerline{
\epsfysize = 2.25truein
\epsfbox[0 0 513 190]
{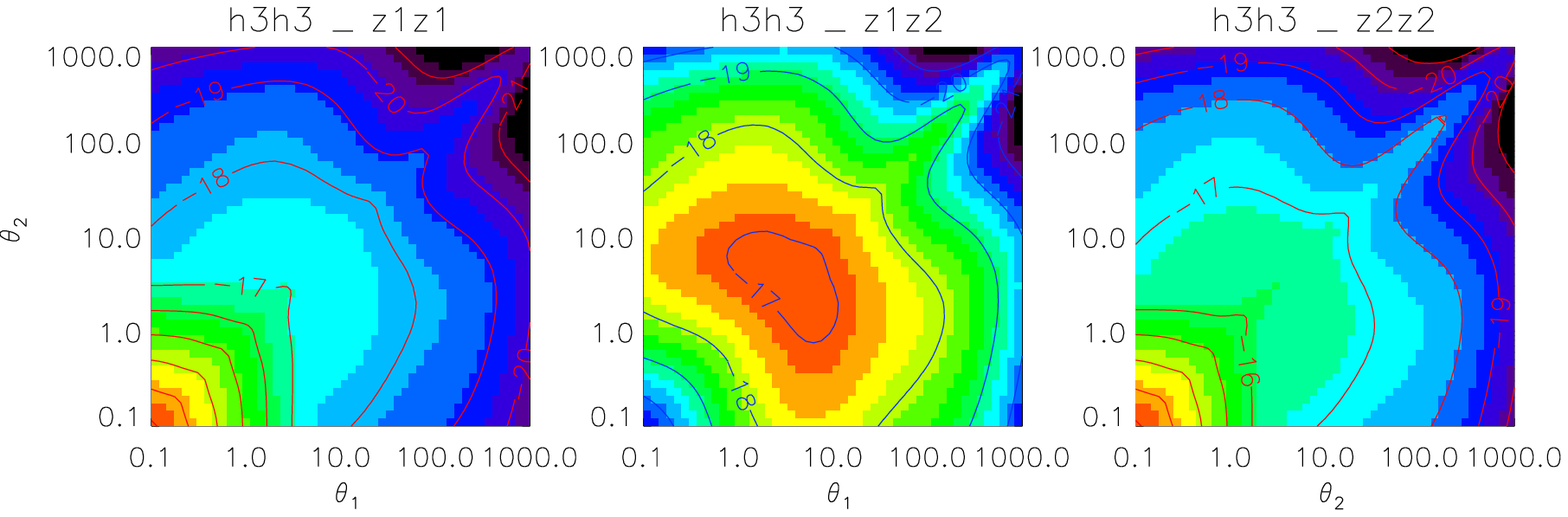}}
\caption{Same as Fig.~\ref{fig:cov_h2h2_zbin} but for $\cH_3$.}
\label{fig:cov_h3h3_zbin}
\end{figure*}

\begin{figure} 
\epsfxsize=8.1 cm \epsfysize=6 cm {\epsfbox{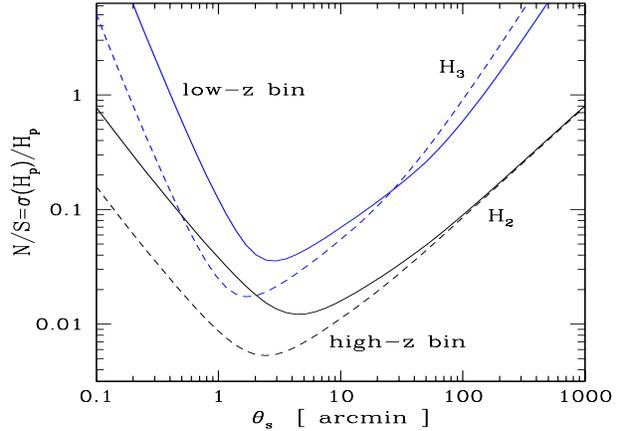}}
\caption{Inverse of the signal to noise ratios of the estimators $\cH_2$ and
$\cH_3$ as in Fig.~\ref{NSH23} but for two redshift bins. The solid lines
correspond to the low redshift sub-sample ($z_s<1.23$) whereas the dashed 
lines show the results for the high redshift bin ($z_s>1.23$).}
\label{NSH23z2}  
\end{figure}

\begin{figure} 
\epsfxsize=8.1 cm \epsfysize=6 cm {\epsfbox{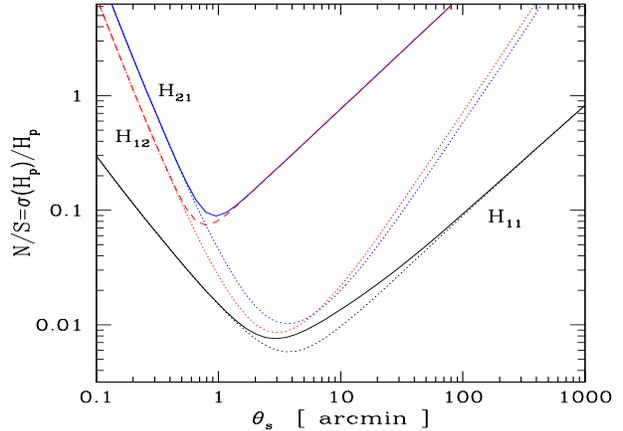}}
\caption{Inverse of the signal to noise ratios of the estimators 
$\cH_{11},\cH_{21}$ and $\cH_{12}$ associated with the cross-correlation 
between both redshift bins. The solid lines are the full noise/signal ratios
while the dotted lines are the results which we obtain when we only keep 
Gaussian terms in the covariance matrices.}
\label{NSH11}  
\end{figure}

We show in Fig.~\ref{NSH23z2} the noise/signal ratio we obtain for each 
redshift bin when we divide the sample into two sub-samples separated by the 
source redshift $z_s=1.23$. The curves are similar to those displayed in
Fig.~\ref{NSH23} for the full sample. For clarity we do not show the curves
associated with the theoretical uncertainties ($f_2$ and $f_3$) or with the
Gaussian approximation to the covariance matrices. They also follow the
behaviour of Fig.~\ref{NSH23}. We can see that the signal to noise ratio is
larger for the high redshift bin (note that we plot its inverse N/S). This
is expected since the longer line of sight leads to a larger amplitude of weak
lensing effects. However at very large angles the signal to noise ratio 
becomes slightly better for the low redshift bin for the estimator $\cH_3$
of non-Gaussianities because of the growth of non-Gaussian gravitational
clustering.

Finally, we show in Fig.~\ref{NSH11} the noise/signal ratios we obtain for
the estimators $\cH_{11},\cH_{21}$ and $\cH_{12}$ which directly measure the
cross-correlation between both redshift bins. Here we defined $\cH_{pq}$
in a manner similar to $\cH_p$ introduced in sect.~\ref{Low-order} such that
their mean is: $\lag\cH_{pq}\rag=\lag M_{{\rm ap}1}^p M_{{\rm ap}2}^q\rag_c$ \
where $M_{{\rm ap}1}$ (resp. $M_{{\rm ap}2}$) is the aperture-mass associated 
with the low (resp. high) redshift bin, see Munshi \& Valageas (2005). 
Of course we can check that the behaviour of different curves is similar to 
that obtained in previous figures.
In particular, the signal to noise ratio is slightly better for $\cH_{12}$ 
than for $\cH_{21}$ since the former gives more weight to the high redshift
bin, in agreement with Fig.~\ref{NSH23z2}. However, we note that the
inclusion of non-Gaussian terms in the covariance matrices now makes a large
difference for the third-order estimators $\cH_{21}$ and $\cH_{12}$.
Indeed, a Gaussian approximation could overestimate the signal to noise ratio 
by up to a factor ten.

The full covariance matrix is displayed in Fig.~\ref{fig:cov_h2h2_zbin}
for $\cH_2$ and Fig.~\ref{fig:cov_h3h3_zbin} for $\cH_3$. In agreement
with the discussion above we find that the covariance is larger for the
high-redshift bin, together with the weak lensing signal itself. As seen
in Fig-\ref{NSH23z2} the overall effect is still to improve the signal to noise
ratio for the gigh-$z$ bin. The shape of the covariance matrix within each
bin (left and right panels) is similar to the behaviour obtained for the
full survey with no redshift binning 
(Figs.~\ref{fig:cov_h2h2},\ref{fig:cov_h3h3}). The cross-correlation between
both bins (middle panels) shows a similarly broad feature due to 
non-Gaussianities. However, there is no rise at small angular scales due to 
shot noise because there are no common galaxies to both bins (hence the shot
noise contribution vanishes).

\section{Estimation of cosmological parameters}
\label{Estimation}

\subsection{Formalism}
\label{Formalism}

Baye's theorem (see e.g. Jaynes, 2003) provides an interesting starting point 
for most parameter estimation studies. Assuming a cosmological data vector 
${\bf a}$ the likelihood function of the cosmological parameter set
${\bf \Theta}$ can be described as:
\beq
{\cal L}({\bf \Theta|a}) = {{{\cal L}({\bf a,\Theta})} {\cal L({\bf \Theta})} \over {{\cal L}({\bf a})}}.
\eeq
Here ${\cal L}(\Theta)$ indicates the prior likelihood function of the 
parameter set $\Theta$ and ${\cal L}(\Theta|{\bf a})$ denotes the conditional 
likelihood function. Normalisation determines $\cal L({\bf a})$ and the prior 
comes from other cosmological observations or as in our case it is assumed to
be constant as we do not include any other observational information. The 
factor ${\cal L}({\bf a}|{\bf \Theta})$ describes the distribution function 
of the observed data vector ${\bf a}$ for a given cosmological
parameter ${\bf \Theta}$. Assuming a multi-variate Gaussian distribution one 
can express ${\cal L}({\bf a}|{\bf \Theta})$ as
\beq
{\cal L}({\bf a}|{\bf \Theta}) = { 1 \over {\sqrt{{(2\pi)}^{\rm N} {\rm det} C(\Theta)}}} {\rm exp}
\left ( -{1 \over 2} {\bf a}^{\rm T} {\bf C^{-1} }{\bf a} \right ),
\eeq
where ${\bf C}^{-1}$ is the inverse of the covariance matrix 
${\bf C}=\lag{\bf a}^T {\bf a} \rag - \lag {\bf a}^T \rag \lag {a} \rag$ of 
the data vector ${\bf a}$ being used. Associated log-likelihood statistic is 
defined as $\chi^2/2$ where $\chi^2 = {\bf a}^{\rm T} {\bf C^{-1}}{\bf a}$.
The covariance matrix ${\bf C}$ is a function of underlying cosmological 
parameters. Its derivatives w.r.t various cosmological parameters along with 
the derivatives of the mean $\mu_{\alpha}$ are computed 
numerically while constructing the Fisher matrix ${\rm F}_{\alpha\beta}$. 

Assuming a fiducial cosmological model, for a given data vector, the 
estimation error of cosmological parameters associated with an unbiased 
estimator can be constructed from the Fisher matrix formalism 
(see Tegmark, 1997, Matsubara \& Szalay 2002, Takada \& Jain 2003).
The Fisher information matrix (Kendall \& Stuart 1969) which is related to 
the the inverse covariance matrix for the cosmological parameter, assesses 
how well the data vector can distinguish the fiducial model from other models.
Thus the error on the parameters $\theta_{\alpha}$ obeys:
$\lag \triangle \theta_{\alpha} \triangle \theta_{\beta} \rag \ge F^{-1}_{\alpha\beta}$.
The left hand side is computed for a specific values of the parameter vector 
$\Theta_0$. The equality is obtained for maximum likelihood estimates of the 
parameters. This inequality - which is also known as Cram\'er-Rao
inequality - provides a minimum variance bound for unbiased estimators and can 
be used to study estimation error and their cross-correlation for various 
parameter sets for a given survey strategy. 
Analytical expression for the Fisher matrix $\rF$ in terms of the covariance 
and mean of the data can be written as (Tegmark et al. 1997):
\beq
\rm F_{\alpha\beta} = \mu_{\alpha} {\rm C}^{-1}\mu_{\beta}+{1 \over 2} Tr \left [ (ln~C)_{\alpha} (ln~C)_{\beta} \right ] 
\eeq
where $\rm (ln~C)_{\alpha} = C^{-1}~C_\alpha$  is the derivative of 
$\rm \ln C$ w.r.t the parameter $\theta_{\alpha}$, and
$\mu_{\alpha}$ denotes the derivative of $\mu = \lag {\bf a} \rag$ w.r.t. 
the parameter $\alpha$. 

The first term corresponds to the case when only mean is being estimated
from the data vector whereas the second term is associated error for variance 
estimations from the data vector ${\bf a}$. The Fisher matrix is a 
positive-definite matrix. It is dominated by the linear order term related 
to estimation of mean of the data. With other cosmological data sets where 
the mean is fixed, such as mean temperature of the CMB, the variance
term can be the dominant term. Here we include both terms in our analysis.
\beq
\rF_{\alpha\beta}(\Theta_0) = \left \langle  {\partial {\rm ln} {\cal L} (a|\Theta) \over \partial \Theta_{\alpha}} 
{\partial {\rm ln} {\cal L} (a|\Theta) \over \partial \Theta_{\beta}} 
\right \rangle
= - \left \langle {\partial^2 {\rm ln} {\cal L} (a|\Theta) \over \partial \Theta_{\alpha}\partial \Theta_{\beta}} \right \rangle
\eeq
Where $\lag \dots \rag$ represents averaging over all possible data 
realization for fixed cosmological parameter values $\Theta=\Theta_0$. The 
equations of ellipsoid defined by the equation 
$\sum_{\alpha,\beta} \Delta \theta_{\alpha} F_{\alpha \beta} \Delta \theta_{\beta} = \lambda^2$ 
define regions in a multidimensional parameter space which can be interpreted 
as error bounds for an experimental setup. A specific choice for $\lambda$ 
defines a $\lambda\sigma$ confidence level. Although strictly speaking such 
an interpretation is valid only when the likelihood function is Gaussian, for 
a mildly non-Gaussian case it can still provide a valuable idea about errors 
associated with estimators.

Therefore, up to a normalisation constant the marginalised two-dimensional 
likelihood function ${\cal L}(\theta_{\alpha},\theta_{\beta})$  for two 
parameters $\theta_{\alpha}$ and $\theta_{\beta}$ can be expressed as:
\beq
 \sim {\rm exp} \left [ -{ 1 \over 2 } (\triangle \theta_{\alpha}, 
\triangle \theta_{\beta})   
\left (
\begin{array} {c c}
{\rF^{-1}}_{\alpha \alpha} & {\rF^{-1}}_{\alpha \beta} \\
{\rF^{-1}}_{\alpha \beta} & {\rF^{-1}}_{\beta \beta}
\end{array} 
\right )
\left (
\begin{array} {c}
{\triangle \theta_{\alpha}}\\
{\triangle \theta_{\beta}}
\end{array}
\right ) 
\right ]
\eeq
Here $({\rm F^{-1}})_{\alpha \beta}$ represents the $\alpha\beta$-element of 
the inverse of the original higher dimensional Fisher matrix $\rm F$. The 
marginalised error-ellipses therefore can be directly deduced from this 
expression. The likelihood contours start to deviate from their Fisher 
counterparts when higher-order correction terms 
start to become important at a large distance from the minima.

In general when a partial set of parameters is being marginalised over the 
other parameters the error covariance of the remaining parameters is given by 
a sub-matrix of the Full inverse Fisher matrix $\rm F^{-1}$. Marginalisation
in general causes a broadening of error ellipses due to reduced level of 
prior information being used.
\beq
\rF = \left(
\begin{array} {c c}
{\rF}_{AA}& {\rF}_{AB} \\
{\rF}_{AB}^{\rm T}& {\rF}_{BB}
\end{array} 
\right)
\eeq
It can be shown that the marginalised $n_A \times n_A$ Fisher matrix 
${\tilde \rF}$ can be expressed as 
${\tilde \rF} = \rF_{AA} - \rF_{BB} \rF_{CC}^{-1}\rF_{BB}^{\rm T} $. The 
second term in this expression provides the correction to the first term, due 
to marginalisation. In the absence of cross-correlations the 
correction term vanishes.

In our study, the data vector ${\bf a}$ consists of various choices of 
measurements of $\cH_2$ and $\cH_3$ at different angular scales for different 
redshift bins ${\bf a}_i = (\cH_2(i;z_a), \cH_3(i;z_a))$. 
Here we have introduced additional index $a$ for
parametrising the redshift bin $z_a$ under consideration. A slightly compact 
notation was used in previous sections where $i$ was allowed to run over 
different angular scales and different redshift bins. Accordingly the 
Covariance matrix has a block structure with various blocks denoting 
covariance of $\cH_2(i;z_a)$ and $\cH_2(i;z_a)$ along with their 
cross-covariances between same and different redshift bins. In our
analysis we have considered various combinations of the estimators $\cH_2$ 
and $\cH_3$ independently and jointly and for one or two redshift bins to 
study errors in parameter estimation. Different cases that
we have considered can be summarised as below.

\begin{itemize}
\item Various angular scales $\theta_s$ with {\it no} redshift information, and only 2-point information from $\cH_2(i)$.
\item Various angular scales $\theta_s$ with {\it no} redshift information, but 2-point $\cH_2(i)$ and 3-point $\cH_3(i)$ information.
\item Various angular scales $\theta_s$ {\it with} redshift information, but only 2-point $\cH_2(i,z_a)$ information.
\item Various angular scales $\theta_s$ {\it with} redshift information, with both 2-point and 3-point 
$\cH_2(i,z_a),\cH_3(i,z_a) $
information.
\end{itemize}

For each of these choices we analyse various combinations of cosmological 
parameters independently or jointly with or without a prior information 
regarding median source redshift $z_0$. Without prior redshift information the 
joint covariance matrix ${\bf C}$ is ill-conditioned and numerical inversions 
are less stable. Due to broad covariance structure of the matrix ${\bf C}$ 
there are no completely independent information at different angular scales 
$\theta_{si}$, especially when there is no redshift information.

\subsection{Use of redshift binning and third-order estimators}
\label{redshift-binning-third-order-estimators}

\begin{figure}
\protect\centerline{
\epsfysize = 1.95truein
\epsfbox[21 413 590 719]
{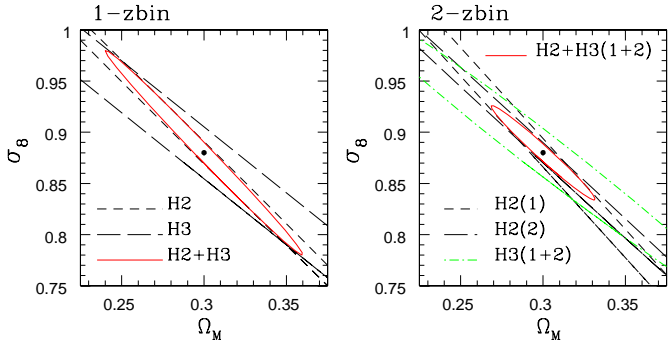}}
\caption{Results of Fisher matrix analysis for the parameter pair 
$\{\Om,\sigma_8\}$, combining the three angular scales $\theta_s=0.1', 10'$
and $1000'$. Perfect knowledge of other parameters is assumed.
For clarity only $3\sigma$ contours are displayed.
{\it Left panel}: Fisher ellipses obtained without redshift binning when
we consider only the variance $\lag\Map^2\rag$ (estimator $\cH_2$, dashed 
line), only the third order cumulant $\lag\Map^3\rag_c$ (estimator $\cH_3$, 
long dashed line), and both cumulants for joint analysis ($\cH_2$ and $\cH_3$, 
solid line).
{\it Right panel}: Fisher analysis with two redshift bins. We consider
the two bins separately for $H_2$ (dashed and long dashed lines) and jointly
for $\cH_3$ only (dot-dashed line) and for the joint pair $\cH_2$ and $\cH_3$ 
(solid line).}
\label{fig:om_sigma_fish1}
\end{figure}

\begin{figure}
\protect\centerline{
\epsfysize = 1.95truein
\epsfbox[21 413 590 719]
{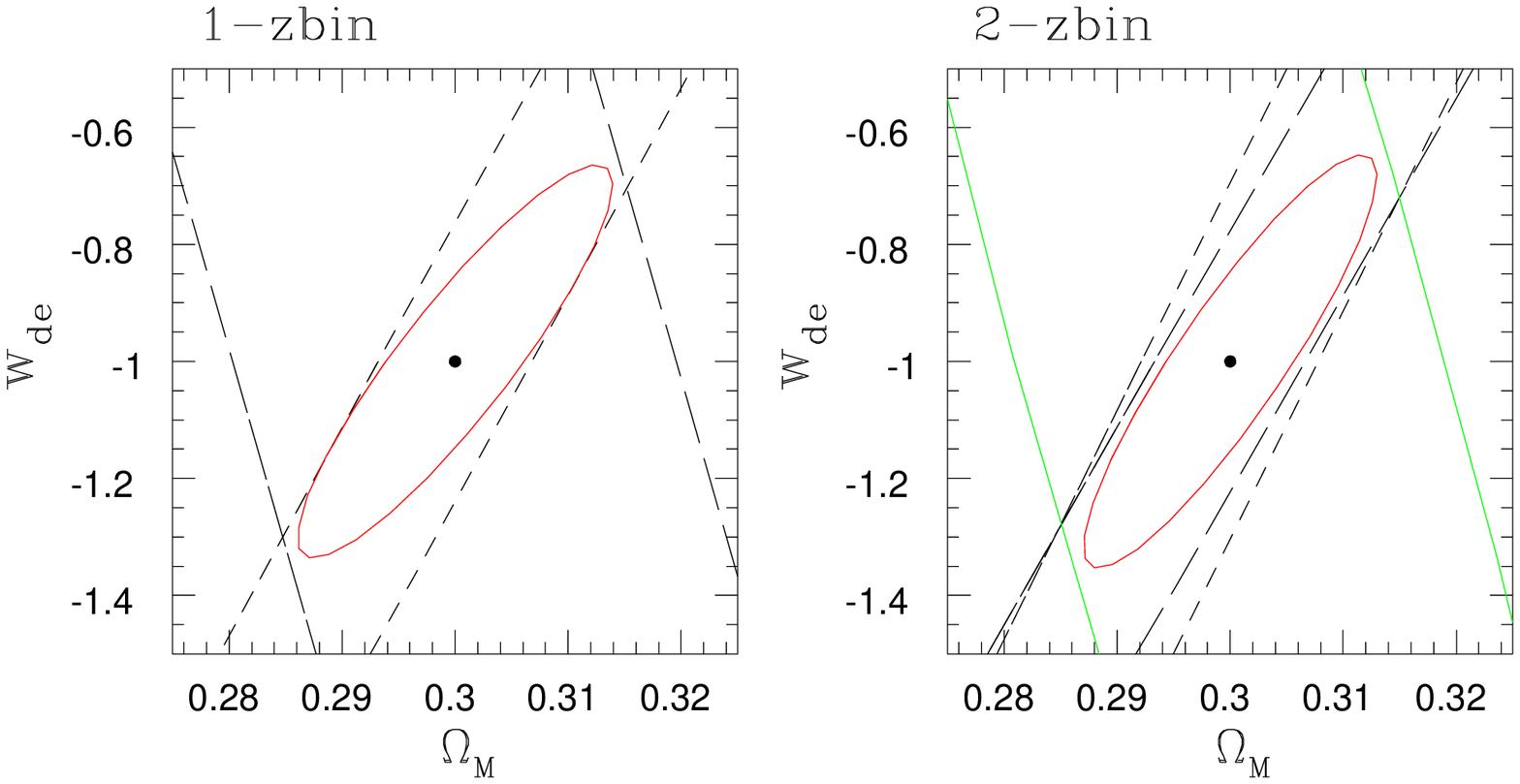}}
\caption{Same as Fig.~\ref{fig:om_sigma_fish1} but for the $\{\Om,\wde\}$ 
pair.}
\label{fig:om_wde_fish1}
\end{figure}

\begin{figure}
\protect\centerline{
\epsfysize = 1.95truein
\epsfbox[21 413 590 719]
{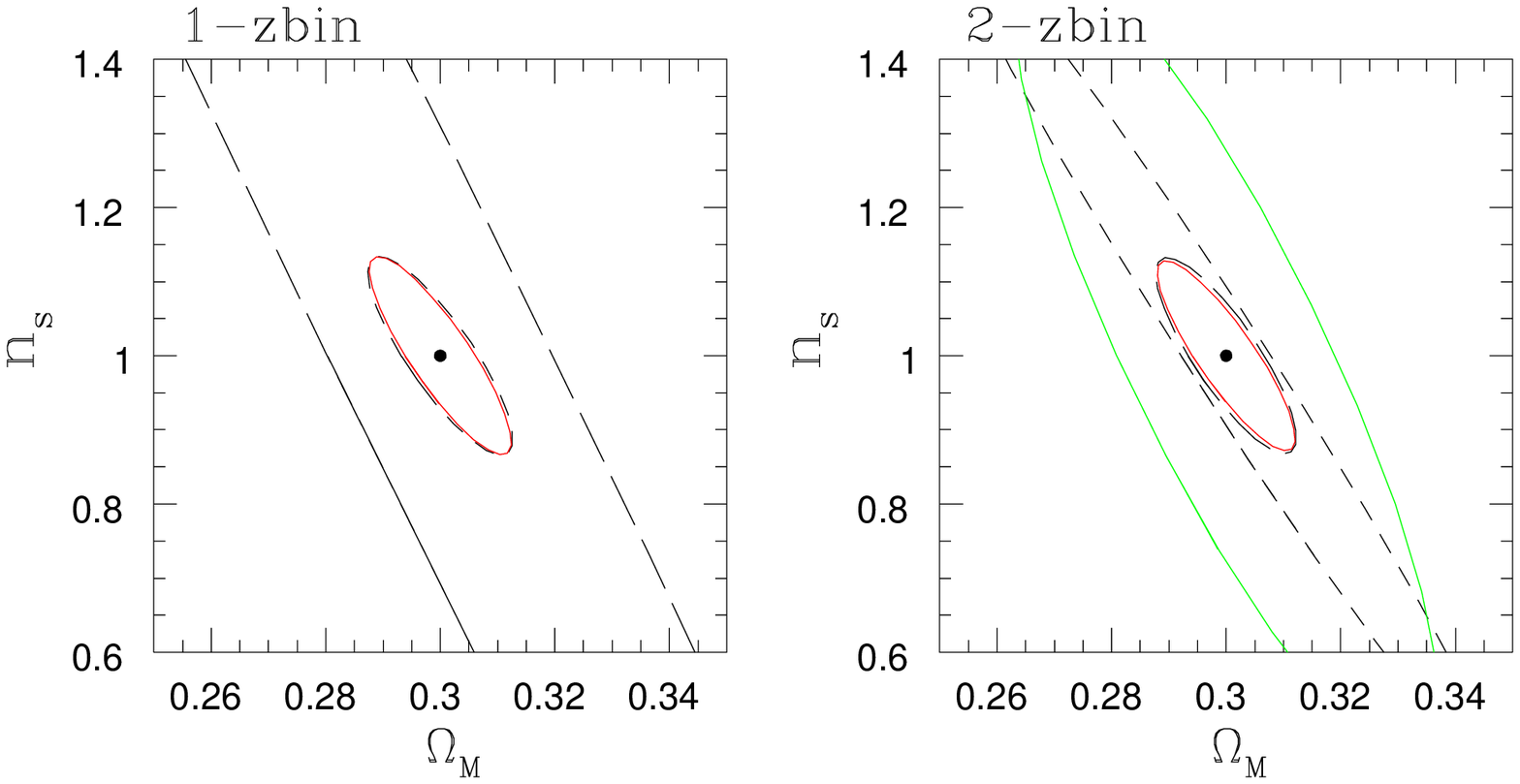}}
\caption{Same as Fig.~\ref{fig:om_sigma_fish1} but for the $\{\Om,n_s\}$ 
pair.}
\label{fig:om_n_fish1}
\end{figure}

\begin{figure}
\protect\centerline{
\epsfysize = 1.95truein
\epsfbox[21 413 590 719]
{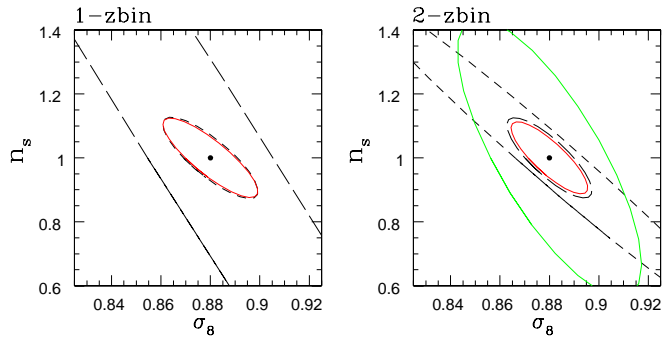}}
\caption{Same as Fig.~\ref{fig:om_sigma_fish1} but for the $\{\sigma_8,n_s\}$ 
pair.}
\label{fig:n_sig8_fish1}
\end{figure}

In Figs.~\ref{fig:om_sigma_fish1}-\ref{fig:n_sig8_fish1} we present
our results from the Fisher analysis described in sect.~\ref{Formalism}
for various cosmological parameter pairs, combining the three angular scales 
$\theta_s=0.1', 10'$ and $1000'$. Perfect knowledge of remaining 
parameters is assumed in each case and only $3\sigma$ contours are plotted 
for clarity. We display both separate and joint analysis of second-order
and third-order cumulants (estimators $\cH_2$ and $\cH_3$) as well as
the results obtained without redshift binning (left panels) and with two
redshift bins (right panels). 
As expected, we find that in all cases the $3\sigma$ contours are much larger 
for the third-order cumulant $\lag\Map^3\rag_c$ than for the variance 
$\lag\Map^2\rag_c$. Higher-order cumulants would be even more noisy.
However, for the pairs $\{\Om,\sigma_8\}$ and $\{\Om,\wde\}$ 
the degeneracy directions (long axis of the ellipse) are
significantly different so that combining $\cH_2$ and $\cH_3$ greatly improves
the constraints on cosmology as compared with $\cH_2$ alone (see left panels). 
Indeed, the long axis of the $\cH_2$-ellipse can be fairly reduced by the 
intersection with the small axis of the $\cH_2$-ellipse. This is most clearly
seen in Fig.~\ref{fig:om_wde_fish1} for the $\{\Om,\wde\}$ pair.
For the $\{\Om,n_s\}$ and $\{\sigma_8,n_s\}$ pairs where the $3\sigma$ 
ellipses associated with $\cH_2$ and $\cH_3$ are almost aligned the
$\cH_2$ contour is well within the $\cH_3$-ellipse so that adding the
third-order cumulant does not tighten the constraints on cosmology.

Next, we show in the right panels the effects of redshift binning.
For the $\{\Om,\sigma_8\}$ pair the $\cH_2$-ellipses have a very high 
eccentricity and are very thin so that the small change of orientation between
the two redshift bins leads to a significant tightening of the intersection
area which improves the constraints on these cosmological parameters.
For other pairs of parameters the $\cH_2$-ellipses associated with the lower
redshift bin (1) are significantly broader than for the high redshift bin and
do not improve the constraints on cosmology. Then, redshift binning is not 
really useful in these cases. The fact that the low redshift bin yields
poorer constraints on cosmological parameters can be understood from the
smaller line of sight which gives less room for weak-lensing effects.

\subsection{Comparison of Fisher analysis and $\chi^2$ likelihood}
\label{Fisherchi2}

\begin{figure}
\protect\centerline{
\epsfysize = 3.25truein
\epsfbox[0 0 370 309]
{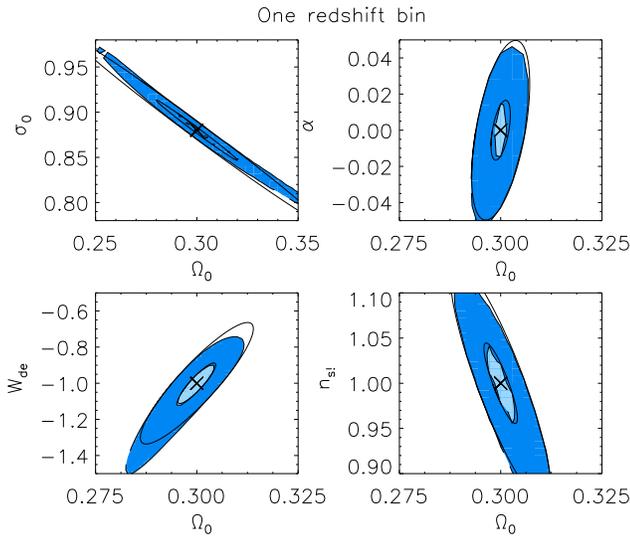}}
\caption{$1\sigma$ and $3\sigma$ Fisher ellipses compared with $\chi^2$ 
contours, combining the three angular scales $\theta_s=0.1',10'$ and $1000'$
without redshift binning. We only display the results corresponding to joint 
analysis of second-order and third-order cumulants.}
\label{fig:chi_1bin_m2m3}
\end{figure}

\begin{figure}
\protect\centerline{
\epsfysize = 3.truein
\epsfbox[0 0 370 339]
{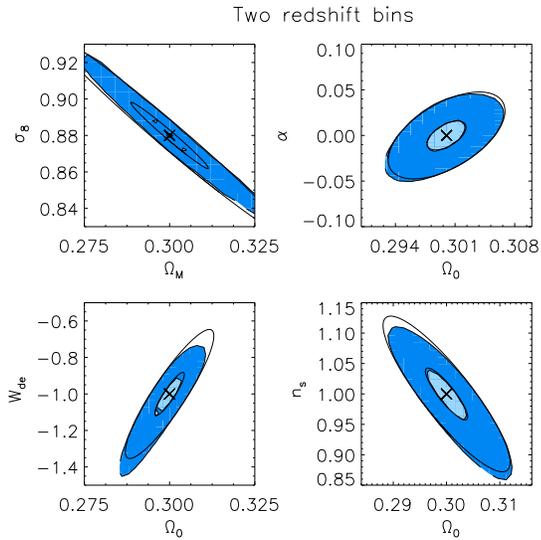}}
\caption{Same as Fig.~\ref{fig:chi_1bin_m2m3} but using tomography. The survey
is split into two redshift bins and their constraints on cosmology are combined
to yield the contours shown in the four panels for various parameter pairs.}
\label{fig:chi_omega}
\end{figure}

\begin{figure}
\protect\centerline{
\epsfysize = 3.truein
\epsfbox[0 0 370 340]
{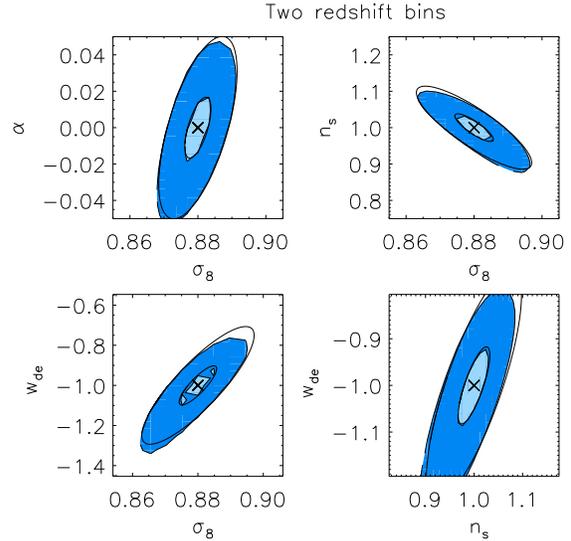}}
\caption{Same as Fig.~\ref{fig:chi_omega} for other cosmological parameter 
pairs.}
\label{fig:chi_sigma}
\end{figure}

We now compare the results of the Fisher matrix analysis with the contour plots
obtained through a $\chi^2$ likelihood function. We again combine the three 
angular scales $\theta_s=0.1',10'$ and $1000'$ and we only display the 
joint analysis of second-order and third-order cumulants. We first show in 
Fig.~\ref{fig:chi_1bin_m2m3} our results with no redshift binning. We can check
that the $\chi^2$ likelihood agrees well with the Fisher matrix analysis. 
Moreover, it happens that both $1\sigma$ and $3\sigma$ contours are close to 
elliptic so that the Fisher matrix analysis appears to be sufficient, except 
for the broader $3\sigma$ contour of the $\{\Om,\wde\}$ pair which is large 
enough to see distortions from the elliptic shape (i.e. far from the fiducial 
model $\{\Om=0.3,\wde=-1\}$ the deviations of cumulants $\lag\Map^p\rag_c$ are 
no longer linear).
Next, we show in Figs.~\ref{fig:chi_omega}-\ref{fig:chi_sigma} a comparison
of Fisher matrix analysis with the $\chi^2$ likelihood function using
redshift binning. In agreement with 
sect.~\ref{redshift-binning-third-order-estimators} we find that tomography
is most efficient for the pair $\{\Om,\sigma_8\}$. Both Fisher matrix and
$\chi^2$ contour plots are consistent, as in Fig.~\ref{fig:chi_1bin_m2m3}.
Again, we find that the $3\sigma$ areas are small enough to have elliptic
shapes.

\subsection{Dependence on uncertainties on the mean redshift}
\label{Mean-redshift}

\begin{figure}
\protect\centerline{
\epsfysize = 3.truein
\epsfbox[0 0 370 330]
{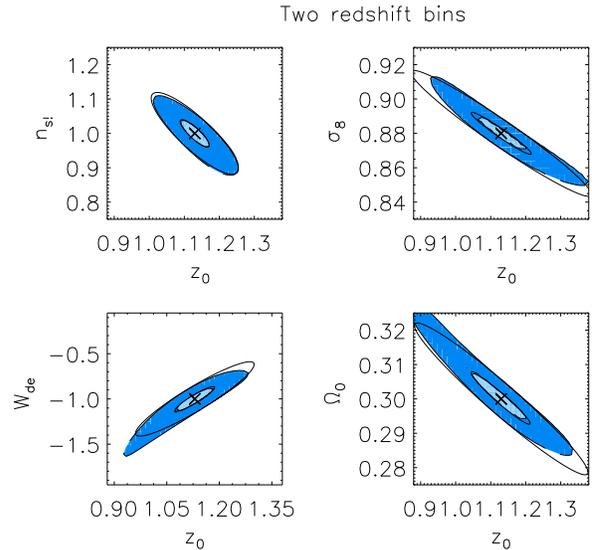}}
\caption{Same as Fig.~\ref{fig:chi_omega} for other cosmological parameter 
pairs including the typical source redshift $z_0$.}
\label{fig:chi_red}
\end{figure}

We now investigate the impact of any uncertainty on the source redshift 
distribution onto weak-lensing results.
Thus, we show in Fig.~\ref{fig:chi_red} Fisher matrix and $\chi^2$ contours
which can be obtained for various cosmological parameters letting the typical
source redshift $z_0$ defined in eq.(\ref{nzSNAP}) free (in each panel other
parameters are assumed to be known). Then, we can directly read in 
Fig.~\ref{fig:chi_red} by how much any inaccuracy on the source redshift
distribution can broaden the constraints on cosmology. Note that even when
we do not impose any a priori on $z_0$ such weak-lensing observations (assuming
again that other parameters are given) are able to recover $z_0$ by themselves
up to $20\%$. If $z_0$ is known to a better accuracy one can cut off the
contours in Fig.~\ref{fig:chi_red}. Depending on the cosmological parameter
of interest we can see that the uncertainty on $z_0$ can multiply the error bar
on the former by a factor of 2 ($n_s$) up to 4 ($\Om$ or $\sigma_8$). 
Therefore, a good accuracy on the source redshifts can be quite rewarding.

\subsection{Marginalizing over unknown parameters} 
\label{Marginalizing}

\begin{figure}
\protect\centerline{
\epsfysize = 3.5truein
\epsfbox[20 146 590 713]
{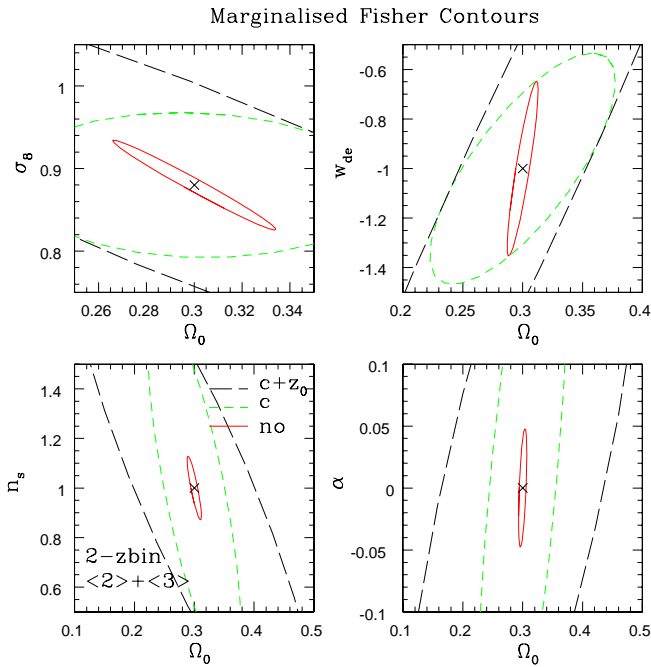}}
\caption{Marginalised 3-$\sigma$ Fisher contours are displayed for various 
pairs of cosmological parameters. The inner solid contours correspond 
to the case when all hidden parameters are assumed to be known perfectly. 
Outermost long-dashed lines correspond to the case when all hidden 
cosmological parameters and median redshift $z_0$ are marginalised. The 
short-dashed line correspond to marginalisation of all hidden cosmological
parameters except $z_0$ (i.e. a perfect knowledge of $z_0$ is assumed). In 
all cases two redshift bins are considered and information at the level of 
two-point and three-point are included for a SNAP class survey. Angular 
scales being considered are $\theta_s=0.1'$, $10'$ and $1000'$ . }
\label{fig:margin_z0}
\end{figure}

Fig.~\ref{fig:margin_z0} shows the individual error ellipses for parameter 
pairs as various degree of marginalisation is used. For the case of two 
redshift bins and joint analysis of 2-point and 3-point analysis we plot 
the effect of having no prior knowledge of $z_0$ or any other parameters on 
estimation error. Clearly accurate knowledge of power-spectrum
and bi-spectrum evolution is essential for putting any constraints on 
spectral index $n_s$ or its run $\alpha_s$. 

Results regarding marginalised errors are presented in units of 
$f_{sky}^{-1/2}$ in table~\ref{table:map1}. Although the entries represent 
an extrapolation of flat-sky results to spherical sky, it is unlikely that 
order of magnitudes are going to change with a more accurate all-sky approach. 
The second column shows $F_{\alpha\alpha}^{-1/2}$ for individual parameters
where complete knowledge of all other parameters is assumed. The third column 
shows marginalised errors when other parameters and $z_0$ are unknown. 
Similarly the fourth column presents errors for additional prior of knowing 
only $z_0$ perfectly. Last column corresponds to the determination of only 
$\Om$, $\sigma_8$ and dark energy equation of state $\wde$ whereas all
other parameters are assumed to be known.

Finally, we present the cross-correlations $r_{ij}$ among estimation errors 
for various parameter pairs in table~\ref{table:cross}. The significant
correlations show that it is difficult to measure simultaneously all 
parameters with a good accuracy, if there are no priors from other data
sets. This is consistent with Fig.~\ref{fig:margin_z0} and with the
comparison between the second and third columns of table~\ref{table:map1}.

\begin{table*}
\begin{center}
\caption{Scatter in estimated parameters in units of ${\rm f^{-1/2}_{sky}}$ 
for individual and joint estimation. Angular scales involved are $0.1'$, $10'$ 
and $1000'$. Values quoted within parenthesis are from two-point analysis 
whereas others are from joint analysis of two- and three-point statistics. 
Information from two redshift bins is considered. Different columns show the
cases where we know all other parameters (2nd column), we marginalize over
other cosmological parameters $X_j$ and redshift $z_0$ (3rd column), we 
marginalize over other parameters $X_j$ only (4th column) and only the three
parameters $\{\Om,\sigma_8,\wde\}$ are unknown.}
\label{table:map1}
\begin{tabular} {@{}lccccc}
\hline
\hline
& $\lag\Map^2\rag$   &  ${\rm F}_{ii}^{-1/2}$ & $[{\rm F}^{-1}]_{ii}^{1/2} \; \{X_j,z_0\} $ & $[{\rm F}^{-1}]_{ii}^{1/2} \; \{X_j\} $ & $[{\rm F}^{-1}]_{ii}^{1/2} \; \{\Om,\sigma_8,\wde\}$ \\
\hline
\hline
& $\sigma(\Om)$ & .00016(.00018)   &  .0059(.037)   & .0022(.0039)  & .0010(.0010)  \\
& $\sigma(\sigma_8)$ & .00025(.00027) & .0075(.088)   & .0025(.0160)  & .0017(.0055)\\
& $\sigma(\wde)$ & .00439(.00626)   & .0606(.523)    & .013(.4425)   & .0116(.1357) \\
& $\sigma(n_s)$ & .00173(.00194)         &  .0305(.082)   & .0207(.0258)  &          -  \\
& $\sigma(\alpha_s)$ &  .00111(.00112)   & .0101(.011)   &  .0083(.0091) &          - \\
& $\sigma(z_0)$ & .00181(.00204)       &  .0370(.405)   &  -     &          - \\
\hline
\hline
\end{tabular}
\end{center}
\end{table*}

\begin{table*}
\begin{center}
\caption{Cross-correlation $r_{ij}$ among estimation errors for various 
parameters. Values quoted within parenthesis are from two-point analysis 
whereas others are from joint analysis of two-point and three-point statistics.
Two redshift bins are considered. Angular scales involved are $0.1'$, $10'$ 
and $1000'$.}
\label{table:cross}
\begin{tabular} {@{}lccccccc}
\hline
\hline
& $r_{ij}$ & $\Om$  &  $\sigma_8$ & $\wde$ &  $n_s$ &  $\alpha_s$ & $z_0$ \\
\hline
\hline
& $\Om$  & $+1.00$   & - &-&-&-&- \\
& $\sigma_8$ & $ -0.88$( $-0.96$) &    $+1.00$ &  &-&-&- \\
& $\wde$ &  $ +0.96$($+0.60$)   & $-0.92$$(-0.37)$   &  $+1.00$  &  &-&- \\
&  $n_s$  &   $-0.87$($-0.97$)   &  $+0.56$$(+ 0.90)$  &  $-0.78$$( -0.63)$  &  $+1.00$  & - & -\\
&  $\alpha_s$ & $ +0.75$($+ 0.59$)   & $-0.37$$(-0.53)$  &  $+ 0.64$$(+0.29)$  &  $-0.97$$(-0.75)$ &   $+1.00$  &  - \\
& $z_0$ & $ +0.92$($+0.99$)   & $-0.94$$(-0.98)$    &  $+0.97$$(+0.53)$   & $-0.73$$(-0.95)$ &   $ +0.58 $$(+0.55)$   &  $+1.00$ \\
\hline
\hline
\end{tabular}
\end{center}
\end{table*}

\section{Discussion}
\label{Discussion}

We have used weak lensing tomography in real space at the level of two-point 
(equivalently power spectrum) and three-point (equivalently bi-spectrum) to 
study how accurately the background dynamics and the nature of dark energy 
can be probed from future weak lensing surveys such as SNAP. It is well known 
that the weight functions and the growth rate dependencies are different 
for the second $\lag\Map^2\rag$ and third-order $\lag\Map^3\rag$ moments.
These complimentary information help to reduce the level of degeneracies 
along with the tomographical information in certain specific choices of 
cosmological parameters.

In our formalism, cross-correlation among various angular scales
at different redshifts can be very easily incorporated in a natural way. Such 
a treatment is completely analytical and is an extension of previous study 
by Munshi \& Coles (2002) which was later extended in Munshi \& Valageas 
(2005). Assuming a ``no-hole'' approach it is possible to directly include 
all contributions to covariance matrices which include cosmic variance at 
large angular scales, shot noise at small angular scales and mixed terms at 
intermediate scales. Taking advantage of a complete analytical description, 
we have check specific approximations to these covariances and their impact 
on error estimates of cosmological parameters. Over-enthusiastic 
simplifications of covariance structure of high-order estimators such as
three-point estimators for $\lag\Map^3\rag$, which depend increasingly on 
accurate description of non-Gaussianities, are typically absent
in harmonic domain based approach and can lead to erroneous error-estimates.

Assuming a WMAP-centric $\Lambda$CDM cosmology, we have considered several
cosmological parameters such as $\Om$, $\sigma_8$, spectral index $n_s$, 
running of spectral index $\alpha_s$, which determine the background dynamics 
of the universe, and the dark energy equation of state $\wde$.

Tomography in some cases can help to reduce the level of degeneracy in 
\{$\sigma_8$, $\Om$\} or \{$\wde$, $\Om$\} however interestingly in a large 
number of other cases most of the information seems to be coming only from 
the higher redshift bin and sub-dividing the sources in different redshift
bins does not seem to affect the estimation accuracy. Equivalently while 
inclusion of higher order moments does indeed help to break the parameter 
degenerecy for some parameter pairs, the other combinations are largely 
unaffected by inclusion of third order information. Clearly both tomography 
and non-Gaussianity information are more useful when a joint estimation of 
several parameters is performed. This is particularly true for spectral index 
$n_s$ and its running $\alpha_s$ which enter the modelling of bi-spectrum 
only through the description of power spectrum. Previous studies of 
tomogrpahy made use of Monte-Carlo simulations of correlated field of views 
(see e.g. Simon et al. 2004) across redshift bins. Our analytical results 
which do not rely on numerical simulations validate and generalise similar 
studies by including detailed descriptions of non-Gaussianities and 
cross-correlatios among various redshift bins.

Generalised third order moments were considered recently by
Kilbinger \& Schneider (2005) which seem to be improving the situation by 
adding extra information at the level of third order. Interestingly however 
the highly correlated nature of such additional input suggests 
that one would only be restricted to a limited range of angular scales for
construction of such a generalised quantities. A more general discussion of 
such generalised statistics which are also named as 
{\it cumulant correlators} in the literature can be found in 
Munshi \& Valageas (2005). In our calculation here we have ignored the 
primordial non-Gaussianity which can potentially be studied
at low redshift given the large sky-coverage of future weak lensing surveys. 
Recent results by Takada \& Jain (2005) showed lensing fields do retain some 
information regarding primordial non-Gaussianity although it is expected that 
most of the information regarding baryon oscillations and primordial 
non-Gaussianities will be erased by large scale non-linearities generated by 
gravitational clustering at a later epoch. Extension of our results in such 
directions to take into account additional cosmological parameters, 
with priors from external data sets such as Type-IA supernova or CMB 
observations, will be presented elsewhere. Additional parameters related to 
time evolution of dark energy equation of state or neutrino mass 
will also be included. Despite large sky-coverage even with future space based 
missions such as SNAP, error of estimation in various cosmological parameters 
remain inherently degenerate due to very nature of observables in weak lensing 
surveys. 
This is particularly underlined by the very high values of degradation factor 
$({\rm F}_{\alpha\alpha})^{-1/2} / ({\rm F}^{-1}_{\alpha\alpha})^{1/2} $ for
various levels of marginalisation as presented in table-\ref{table:map1}. In 
table-\ref{table:cross} scaled version of inverse Fisher matrix 
$r_{\alpha\beta}$  which is also the cross correlation coefficient 
$r_{\alpha\beta} = ({\rm F}^{-1})_{\alpha\beta} / \{({\rm F}^{-1})_{\alpha\alpha}({\rm F}^{-1})_{\beta\beta}\}^{1/2}$ 
of estimation error of different parameters is plotted for certain choices of 
the parameters as indicated. Our findings are in agreement with a recent study 
by Kilbinger \& Munshi (2005) where they reached a similar conclusion based
on more elaborate generalised eigenmode analysis of Fisher matrices.

% info about z_0

Knowledge of redshift distribution of source galaxies from external 
observations can help parameter estimation especially for smaller surveys. 
Even for larger surveys such as SNAP we found that order of magnitude 
improvement in accuracy is possible for almost all parameters we have studied.
Large scale spectroscopy of a fair sample of faint source galaxies may not be 
possible, alternatively photometric redshift determination of subsample of 
observed galaxies can still improve the level of accuracy. It is likely that 
future observations with near complete sky coverage $f_{sky} \sim 1$ spanning 
large redshift coverage can deal with much more detailed description of the
source redshift distributions which we have only parametrised by one parameter 
$z_0$.

% other cosmo observations

Several authors have recently studied joint constraints on various 
cosmological parameters such as $\{n_s,\sigma_8\}$ and $\{\alpha_s, n_s\}$ 
using Lyman-alpha forest data jointly with 1-year observations from WMAP 
(Spergel et al. 2003). Clearly the future space based SNAP class experiments 
will help to provide competitive constraints when used jointly with external 
data set.

%importance of knowledge of f_2 and f_3

Bi-spectrum of density fluctuations carries complementary information 
regarding background geometry and dynamics of the universe. Although current 
surveys are at the detection limit (see Bernardeu 2002) of measuring 
significant non-Gaussianity, it is expected that future space based 
observations with much greater sky coverage will supplement power spectrum 
information with accurate measurement of bi-spectrum. As we have shown here, 
this will require a very accurate description of evolution of bi-spectrum. 
Associating unknown parameters such as $f_2$ and $f_3$ with low order 
description of dark matter clustering and trying to estimate them directly 
from the data degrades the estimation accuracy of other cosmological
parameters considerably. 

%Fisher vs. MCMC

Finally, in our studies we have compared Fisher analysis with full grid
based $\chi^2$ calculations. Importantly Fisher analysis should only be seen 
as a way to estimate minimum variance errors for unbiased estimators and to 
check degeneracy directions. It does not propose to reveal the detailed 
behaviour of the confidence plane far from the fiducial models.
This is particularly true for very asymmetric constraints e.g. in 
$(\Om,\sigma_8)$ plane. Nevertheless the various parameter pairs that we have 
studied show a reasonable match in confidence level at $1\sigma$ and even to 
$3\sigma$ levels. An alternative approach which uses entropy functional
to map out the likelihood distribution in the parameter space was considered 
in Taylor \& Watts (2002). A grid based calculation of full $\chi^2$ becomes
prohibitively costly as the dimension of the parameter space increase. A more 
effective sampling of the parameter space can be achieved by Monte Carlo based 
Markov-Chain algorithms. A full implementation will be presented elsewhere. 

% other surveys

In our present analysis we have focused on survey parameters similar to 
proposed SANP-class surveys (JDEM). It is not difficult to extend it to other 
survey specifications. With future well optimised surveys with large sky 
coverage, CFHTLS (172 deg$^2$), SNAP (300 deg$^2$), VISTA (10000 deg$^2$), 
Pan-starr (31000 deg$^2$) and recent data analysis techniques,
weak lensing will be a very useful tool to probe not only standard 
cosmological parameters, such as $\Om$ and $\sigma_8$, but also the dark 
energy equation of state parameters such as $\wde$. However to achieve this 
goal it is essential to have a tight handle on systematics which should at 
least stay comparable to statistical errors.

%\begin{figure*}
%\protect\centerline{
%\epsfysize = 1.9truein
%\epsfbox[0 0 420 215]
%{comp_approx.eps}}
%\caption{3$\sigma$  $\chi^2$ contours are plotted for various approximations used. Two different
%parameter pairs are chosen as an example. Angular scales used are $\theta_s=.1'$, $\theta_s = 30'$
%and $\theta_s=100'$.}
%\label{fig:aprrox}
%\end{figure*}

\section*{acknowledgments}
DM was supported by PPARC of grant
RG28936. It is a pleasure for DM to acknowledge many fruitful discussons with
memebers of Cambridge Planck Analysis Center (CPAC). DM also thanks  Martin 
Kilbinger, Alan Heavens, Yun Wang, Lindsay King and George Efstathiou for useful discussions.

\appendix

\section{Other survey characteristics}
\label{Other-survey-characteristics}

\begin{figure} 
\epsfxsize=8.1 cm \epsfysize=6 cm {\epsfbox{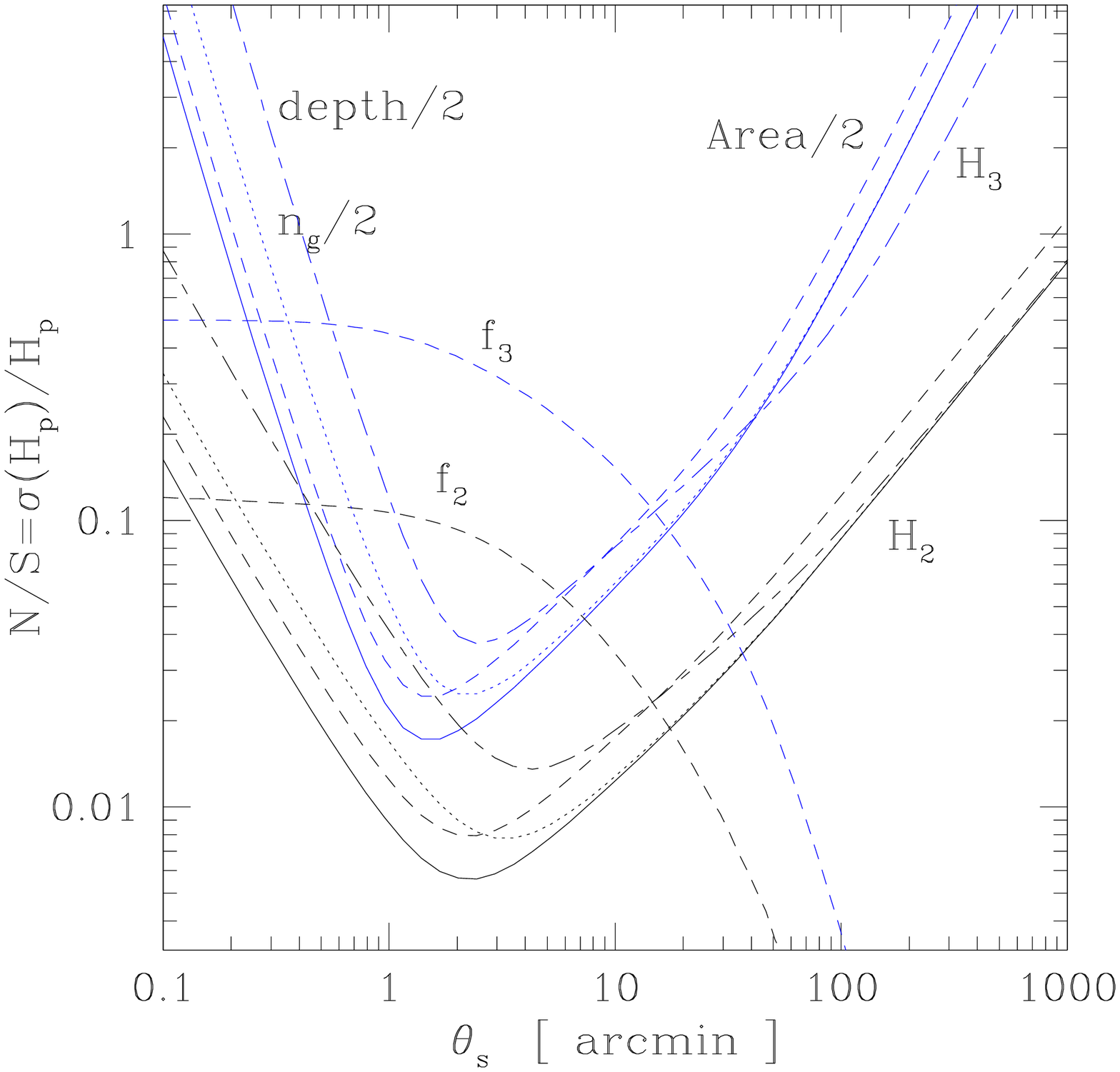}}
\caption{Inverse of the signal to noise ratios of the estimators $\cH_2$ and
$\cH_3$ of the second-order and third-order cumulants of the aperture-mass
for various survey properties. The solid lines correspond to the SNAP survey
as in Fig.~\ref{NSH23}, the dashed lines to a similar survey with an area twice
smaller ($A=150 {\rm deg}^2$), the dotted lines to a galaxy number density
twice smaller ($n_g=50 {\rm arcmin}^{-2}$) and the dot-dashed lines to
a survey with a depth which is twice smaller ($z_0=0.57$ and 
$z_{\rm max}=1.5$). There is no redshift binning and we use the full covariance
$C_{ii}$.}
\label{NSH23all}  
\end{figure}

\begin{figure} 
\epsfxsize=8.1 cm \epsfysize=6 cm {\epsfbox{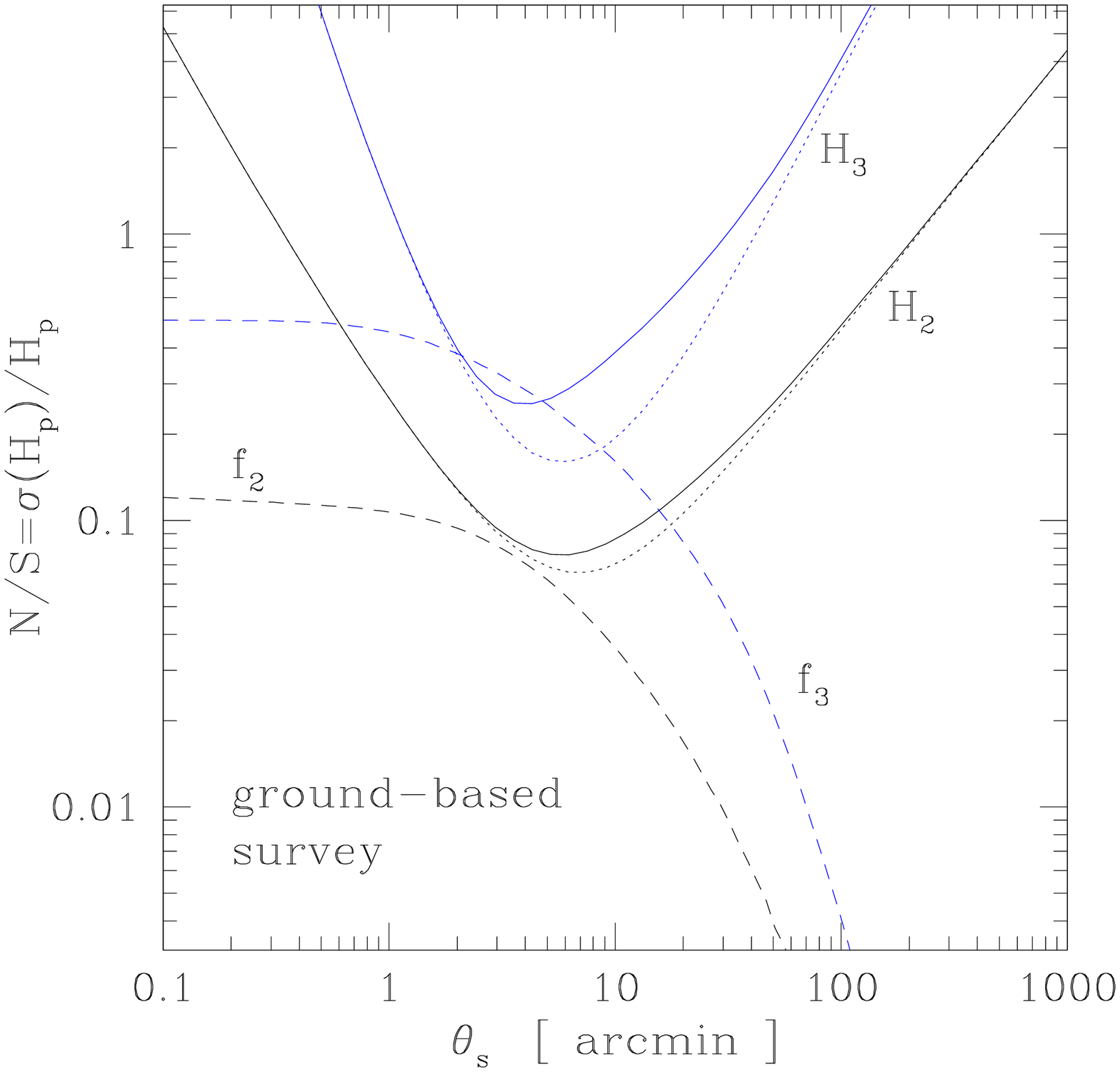}}
\caption{Inverse of the signal to noise ratios of the estimators $\cH_2$ and
$\cH_3$ of the second-order and third-order cumulants of the aperture-mass
as in Fig.~\ref{NSH23} but for a typical ground-based survey. There is no 
redshift binning. The solid lines correspond to the full covariance $C$
while the dotted line show the results we obtain when we only use its Gaussian
part $C^G$.}
\label{NSH23ground}  
\end{figure}

For completeness, we display in this appendix the noise to signal ratios we
obtain for different survey properties. Thus, we display in Fig.~\ref{NSH23all}
our results for the SNAP survey as in Fig.~\ref{NSH23} (solid lines) as well 
as for the cases when we divide by a factor two its area (dashed line), its 
galaxy number density (dotted lines) or its depth (dot-dashed lines).
When the depth of the survey is varied the dashed curves labeled $f_2$ and
$f_3$ change slightly (as the redshift distribution of sources whence of
the density fluctuations probed by the survey is varied) but since the
modification is very small (by definition the curves show the same plateau at
small scales) we do not add them in Fig.~\ref{NSH23all} for clarity. 
We can check that a smaller galaxy number density increases the noise at small 
scales (below $2'$) which is dominated by shot noise but does not change the 
signal to noise ratio at large scales which is governed by the cosmic 
variance. By contrast, a smaller area leads as expected to a smaller signal 
to noise ratio at all scales. Finally, a smaller depth decreases the signal 
to noise ratio at small scales as the amplitude of gravitational lensing 
effects is weaker because of the smaller line of sight while the shot noise
remains the same. At large scales the signal to noise ratio for $\cH_2$
is not diminished while it improves somewhat for $\cH_3$ (at fixed galaxy
number density) as the survey focuses more on the non-linear regime
of gravitational clustering where non-Gaussianities are larger.

Next, we show in Fig.~\ref{NSH23ground} the noise to signal ratios obtained
for a typical ground-based survey with a galaxy distribution $n(z_s) \propto
z_s^2 e^{-(z_s/0.8)^{1.5}}$, an area $A=10$ deg$^2$ and a galaxy number 
density $n_g=20$ arcmin$^{-2}$ (e.g., van Waerbeke et al. (2001)).  
Of course, we can check that both the shot-noise and the cosmic variance
are significantly larger than for a space-based mission such as SNAP.
This implies that the range of scales which can be used to extract cosmological
information is quite smaller. On the other hand, the theoretical uncertainty
on the non-linear regime of gravitational clustering (denoted by the dashed
curves) is only of the same order as other sources of noise around $5'$ and
is relatively smaller at other scales, so that it is not really the limiting 
factor for such surveys.  

A generalised eigen mode analysis (Lo\'eve 1948, Karhunen 1947) which can address 
the issue of optimisation of
survey design $(z_0,f_{sky},n_g)$ for recovery of a given set of cosmological
parameters in line with Kilbinger \& Munshi (2005) but including tomography and
non-Gaussianity will be presented elsewhere.

\end{document}